\documentclass{trbunofficial}
\usepackage{graphicx}

\usepackage[hidelinks]{hyperref}
\usepackage{amsmath}
\usepackage{amsfonts}
\usepackage{caption}
\usepackage{subcaption}
\usepackage{hyperref}

\AuthorHeaders{Hou, Kusari, and Lin}
\title{Evaluating Parametric Car-Following Models in Naturalistic Congestion: Insights in Driver Behavior and Model Limitations}

\author{%
  \textbf{Huaidian Hou*, Corresponding Author}\\
  houhd@umich.edu\\
  \hfill\break%
  \textbf{Arpan Kusari, Ph.D.}\\
  kusari@umich.edu\\
  \hfill\break%
  \textbf{Brian T. W. Lin, Ph.D.}\\
  btwlin@umich.edu
}

\TotalWords{6509}

\begin{document}
\maketitle

\section{Abstract}

Car-Following is a broadly studied state of driving, and many modeling approaches through various heuristics and engineering methods have been proposed. Congestion is a common traffic phenomenon also widely investigated, both from macroscopic and microscopic perspectives. Yet, current literature lack a unified evaluation of Car-Following models with naturalistic congestion data. This paper compares the performance of five parametric Car-Following models: IDM, ACC, Gipps, OVM, and FVDM, using a rich naturalistic congestion dataset. The five models in question is found to perform similarly when optimized over the same RMSNE metric. Sub-sequences of Car-Following where models noticeably disagree with driver behavior is noticed and separately investigated. A review of corresponding front-facing and cabin video data reveals distraction and driving with momentum as potential reasons for model-reality difference. We further show that drivers often employ coasting and idle creep under Car-Following in different speed ranges, which existing parametric models fail to capture. Finally, time-series clustering is performed and analysis of result clusters align with empirical findings. 

Our findings highlight the necessity to consider vehicle dynamical properties including coasting and idle creep abilities, which drivers take extensive use of under low speed congestions. Future research could integrate such parameters with traditional parametric models to improve congestion modeling performance. We also suggest future research into investigating temporal correlations between clustered blocks to reveal behavioral transition patterns exhibited by drivers in congestions. Source code for this study can be found on \href{https://github.com/DanielHou315/Car_Following_Eval}{Github}. 

\hfill\break%
\noindent\textit{Keywords}: Car-Following, Congestion, Coasting and Idle Creep, Time-Series Clustering
\newpage

\section{Introduction}

Car-Following (CF) is a fundamental component of driving behaviors, where drivers' behaviors are primarily affected by the trajectory of the lead vehicle (LV). Car-Following behavior have been extensively studied since the 1950s, and various parametric car following models, aimed at describing driver behaviors in Car-Following events, have been proposed. These Car-Following models are a class of prediction models that predicts one aspect of driving maneuver (such as acceleration and braking) using various heuristics. The development of digital assistive technologies since early 2000s enabled vehicles to automate Car-Following behaviors with Adaptive Cruise Control. Such features leverage the aforementioned prediction models to control the ego-vehicle according to external sensor measurements. Recent development in Vehicle-to-Vehicle and Vehicle-to-Everything technologies also bears potential of improving Car-Following behavior by introducing more context to the driving-behavior prediction. The collection of large-scale real-world driving datasets, where human driving behavior is not affected by explicit experimental configurations, provides platforms to study natural driver beahviors and reactions in a real-world setting. 

While parametric Car-Following models have been widely studied, to our knowledge, limited literature exists in studying how well these models predict human behaviors in low-speed congestion scenarios with real-world datasets. This study aims to investigate the performance of selected parametric Car-Following models, which performs well in high-speed free flow traffic, in low-speed congested situations. Specifically, we investigated the following main questions: 

\begin{enumerate}
    \item Do parametric Car-Following model predictions agree with human behavior in congested situations as they are in highways, using highway-optimized parameters?
    
    \item Do parametric Car-Following model predictions agree with human behavior in congested situations using congestion-optimized parameters?
    
    \item In what congestion scenarios do parametric Car-Following models fail to capture human behaviors?

    \item Where parametric models fail to explain driver behaviors, what actions are performed differently by the drivers and why?
\end{enumerate} 
In the next sections of this paper, we evaluated model performance through traffic simulation on real-world congestion data. We also show that driver anticipation and distraction associates with large model-reality difference. Furthermore, the model-reality gap analysis reveals two vehicle dynamics features drivers take advantage of in real-world congestions that are currently ignored by kinematic Car-Following models. 

\section{Literature Review}

\subsubsection{Car-Following Models}

Car-Following models have been widely studied since early 1900s, and models with different heuristics have been proposed over the last century. Notably, kinematics, human psychology, and control theory are amongst the primary heuristics for parametric Car-Following model designs. Kinematics-based models are a popular genre of parametric Car-Following models for their robustness, simplicity and interpret-ability. In 1958, \citep{chandler_traffic_1958} propose the GHR model as a stimulus-response model, which provides a generalized analysis framework for future stimulus-based modeling of Car-Following. \citep{gipps_queueing_1977} proposed the Gipps model as a safe-distance based model, which models drivers reaction as intention to maintain a safe stopping distance to the lead vehicle. \citep{treiber_congested_2000} proposed the Intelligent Drive Model (IDM) to unify both optimal velocity and optimal distance as one continuous function. \citep{al-jameel_examining_2009} analyzed the asymptotic limitations of IDM and proposed various improvement techniques. \citep{kesting_enhanced_2010} addresses limitations of IDM by introducing Constant Acceleration Heuristic to form the ACC model. \citep{bando_dynamical_1995} proposed the Optimal Velocity family of model, which views Car-Following as attempts to always reach optimal velocity under specific circumstances. The behavior of this model with explicit delay is studied in \cite{bando_analysis_1998}. Built upon OVM, \citep{jiang_full_2001} proposed the Full Velocity Difference Model (FVDM) to include the difference in velocity as a factor of acceleration output. Several studies compared the performance of kinematic Car-Following models on real-world traffic datasets. \citep{zhang_comprehensive_2021} compared GHR, IDM, Gipps, and Wiedemann models in mixed road conditions in Shanghai and reveals error distributions for each model under three driving styles. \citep{kim_identifying_2023} Compared IDM, ACC, and MIXIC models on naturalistic highway driving data, and the model trajectories are investigated.  

Additionally, psychology-based models are developed to model aspects of human cognition and decision making processes. Action Point Models (APM) aim to model human perception sensitivity and response bias when determining acceleration/brake timing \cite{zhang_car-following_2024}. The model proposed by \citep{wiedemann_simulation_1974}, used in VISSIM, models Car-Following with four different regimes along with their respective boundary functions. Control theory based Car-Following models also gained much attention for their robustness and broad potential in industrial applications. All of Linear, Nonlinear, and MPC based controllers have been explored in the context of Car-Following \cite{zhang_car-following_2024}. 

Naturalistic dataset for Car-Following is crucial in evaluating theories of human behavior in a real-world setting. The HighD dataset by \citep{krajewski_highd_2018}, composed of data collected on German highways, supports the study of complex multi-vehicle interaction through birds-eye-view data. The NDS dataset \cite{campbell_safety_2012} contains extensive and diverse driving data from 3500 drivers around 6 locations in the United States. \citep{chen_follownet_2023} proposes FollowNet, which integrates the works of HighD, NGSIM, SPMD, Waymo, and Lyft into a unified interface to enable larger-scale studies and benchmarking. The IVBSS project \cite{sayer_integrated_2011}, fulfilled by the University of Michigan Transportation Research Institute (UMTRI), results in a rich set of naturalistic driving history data as a byproduct of evaluating a novel vehicle safety system. 

\section{Definition of Congestion} \label{def-of-congestion}

In \cite{kerner_traffic_2009}, traffic congestion is defined as a phase of traffic where the speed is reduced sharply and traffic density increases in an initially free traffic flow. We adhere to this definition and observe that most congestions are accompanied by, on a microscopic level, frequent acceleration and deceleration by a driver. Therefore, we identify congestion as an extended period of a trip where frequent acceleration and braking behavior is observed, primarily as reaction to slow-moving lead vehicles. These acceleration and braking are otherwise unnecessary should the driving take place in free flow. This definition enables us to extract congestion events from data recordings available in a personal vehicle, rather than needing global information of the traffic situation. 

Still, there exists situations other than congestion that could cause the above phenomenon. A combination of traffic lights and crossing pedestrians in urban road, for example, lead to frequent and unnecessary acceleration. Lacking such input, the traditional parametric Car-Following models are unable to capture such complex vehicle-to-environment interaction. To study what drivers do specifically in congestions, we focus on the congestion events on highways, where interactions with road traffic participants other than the lead vehicle are fairly limited. Thus, we propose the following microscopic criteria for identifying continuous congestion events: 

\begin{itemize}
    \item Driver frequently applies the brakes, at least once every T seconds. We found $T = 90$ to cover the vast majority of congestion scenarios. 
    \item Driver drives slowly on average between brakes with $\bar{v} \leq v_\text{max}$. In this study, we use a threshold of $v_\text{max} = 40$ km/h as the threshold of mean speed between two brakes. 
    \item The road type to be highway in at least 50\% of the time segment identified. 
\end{itemize}
The above criterion is used to create the congestion dataset used throughout the rest of the study.

\section{Methods}

\subsection{Candidate Models}

For this study, we selected the Gipps model, the Intelligent Driver Model (IDM), the Adaptive Cruise Control (ACC) model with Improved IDM \cite{treiber_traffic_2013}, the Optimal Velocity Model (OVM), and Full Velocity Difference Model (FVDM). The selection of models aims to both include different heuristics and base-improvement pairs models to compare their characteristics and efficacy. 

\subsubsection{Gipps Model}
The Gipps model is considered the most commonly used parametric model with the safe-distance (or collision avoidance) heuristic \cite{vasconcelos_calibration_2014}. It predicts the velocity of the ego vehicle using acceleration and deceleration sub-modules to balance between safe following distance and desirable speed. The Gipps model also explicitly considers a response time (delay) factor, whereas other models investigated in this study assumed no response time. The speed output of the Gipps model can be expressed as 

$$
    v(t+\tau) = \min(v^{\text{acc}}(t+\tau), v^{\text{dec}}(t+\tau))
$$ \\
and each of the components is expressed as

$$
    v^{\text{acc}}(t + \tau) = v(t) + 2.5 a_{\text{max}} \tau \left( 1 - \frac{v(t)}{v_{\text{opt}}} \right) \sqrt{0.025 + \frac{v(t)}{v_{\text{opt}}}}
$$
$$
    v^{\text{dec}}(t + \tau) = -\tau b_{\text{max}} + \sqrt{\tau^2 b_{\text{max}}^2 + b_{\text{max}} \left\{ 2 \left[ s(t) - s_0 \right] - \tau v(t) + \frac{(v_{\text{lv}}(t))^2}{b_{\text{max}}^{\text{lv}}} \right\}}
$$ \\
where \begin{itemize}
    \item $\tau$ is the reaction time in $s$
    \item $v_{\text{opt}}$ is the desired velocity in $m/s$
    \item $a_{\text{max}}$  is the desired acceleration in $m/s^2$
    \item $b_{\text{max}}$ is the desired deceleration in (negative) $m/s^2$
    \item $s_0$ is the minimum safe distance in $m$
    \item $b^{\text{lv}}_{\text{max}}$ is the estimated maximum braking power of the lead vehicle in (negative) $m/s^2$
\end{itemize}

\subsubsection{IDM}
The Intelligent Driver Model (IDM) proposed by \citep{treiber_congested_2000}. IDM and its variants are recognized as more accurate and robust models in \cite{zhang_comprehensive_2021}, \cite{kim_identifying_2023} with comparable number of tune-able parameters. The acceleration produced by IDM is calculated as

$$
    a_{\text{idm}}(t) = a_{\text{max}} \left[ 1 - \left( \frac{v(t)}{v_{\text{opt}}} \right)^4 - \left( \frac{s_{\text{opt}}(t)}{s(t)} \right)^2 \right]
$$ \\
where $a_{\text{max}}$ is the maximum acceleration in $m/s^2$, $v_{\text{opt}}$ is the desired velocity and $s_{\text{opt}}(t)$ is the desired space headway, computed as

$$
    s_{\text{opt}}(t) = s_0 + \max \left( 0, v(t)T - \frac{v(t)\Delta s(t)}{2 \sqrt{a_{\text{max}} b_{\text{comf}}}} \right)
$$\\
where \begin{itemize}
    \item $b_{\text{comf}}$ is the comfortable deceleration in (negative) $m/s^2$
    \item $s_0$ is the minimum safe distance in $m$
    \item $T$ is the desired time headway in $s$
\end{itemize}

\subsubsection{ACC}
ACC is an adaptation of IDM proposed by Kesting et al. \cite{kesting_enhanced_2010} to include the Constant Acceleration Heuristic (CAH). For this particular ACC, we use the Improved IDM (IIDM) described in \cite{treiber_traffic_2013} which is compared alongside IDM in \cite{kim_identifying_2023}. The ACC model takes into account two accelerations produced by separate heuristics: IIDM and CAH: 

$$
    a_{acc}(t) = \begin{cases}
        a_{\text{iidm}}(t) &\text{if } a_{\text{iidm}}(t) < a_{\text{cah}}(t) \\ 
        (1-c)a_{\text{iidm}}(t) + c (a_{\text{cah}}(t) + b_{\text{comf}} \cdot  \tanh(\frac{a_{\text{iidm}}(t) - a_{\text{cah}}(t)}{b_{\text{comf}}})) &\text{otherwise}
    \end{cases}
$$ \\
where $c$ is a constant factor used to combine two accelerations in the second regime. Here, we use $c=0.99$ following the original practice of \citep{kesting_enhanced_2010}. The IIDM acceleration $a_{\text{iidm}}$ is calculated in a per-regime basis given as 

$$  
    a_{\text{iidm}} = \begin{cases}
        a_\text{max}(1-z^2) \quad &\text{if } v(t) \geq v_0, z \geq 1 \\
        a_{\text{free}} (1-z^{2a/a_{\text{free}}}) &\text{if } v(t) \geq v_0, z < 1 \\
        a_{\text{free}} + a(1-z^2) &\text{if } v(t) < v_0, z \geq 1 \\
        a_{\text{free}} & \text{otherwise}
    \end{cases}
$$ \\
where $z = \frac{s_{\text{opt}}(t)}{s(t)}$ with $s_{\text{opt}}(t)$ from IDM and 

$$
    a_{\text{free}} = \begin{cases}
        a_\text{max} \left[1-(\frac{v(t)}{v_0})^\delta \right] &\text{if } v(t) \leq v_0 \\
        b_\text{comf} \left[1-(\frac{v_0}{v(t)})^\frac{a_{\text{max}}\delta}{b_{\text{comf}}} \right] &\text{otherwise}
    \end{cases}
$$ \\
where $\delta$ is an exponent constant, usually set to 4 according to\cite{kim_identifying_2023} as we do throughout this study. 
Separately, in a given situation, the maximum crash-free acceleration is given as

$$
    a_{cah} = \begin{cases}
        \frac{v(t)^2 \tilde{a_{\text{lv}}}}{v(t)^2 - 2s(t)\tilde{a_{\text{lv}}}} \quad &\text{if } v_{\text{lv}}(t) \cdot \Delta s(t) \leq 2s(t) \cdot \tilde{a_{\text{lv}}} \\
        \tilde{a_{\text{lv}}} - \frac{(\Delta s(t))^2 [[\Delta s(t) \geq 0]]}{2s(t)} \quad &\text{otherwise}
    \end{cases}
$$ \\
where $\tilde{a_{\text{lv}}} = min(a_{lv}(t), a_{max})$ is used to resolve artifacts in situations where the lead vehicle has a larger acceleration as explained in \cite{kesting_enhanced_2010}. 

\subsubsection{OVM}
The Optimal Velocity Model (OVM), originally proposed by \citep{bando_dynamical_1995} represents the family of models using the optimal velocity heuristics. This model predicts the acceleration by considering the difference between optimal velocity $\tilde{V}(t)$ and actual velocity $v(t)$ as a stimulus. The OVM acceleration is expressed as

$$
    a(t) = \alpha \left[ \tilde{V}(t) - v(t) \right]
$$ \\
where $\alpha$ is the stimulus-response factor of the difference between optimal and actual velocity. The optimal velocity $\tilde{V}$ is produced by the Optimal Velocity Function (OVF) that is usually a $\tanh$ function of space gap. This function models the driver's preference to drive at low speeds when gap is small, and the desired velocity increases to a maximum as the gap enlarges. Various modifications of the OVF based on the original function have been proposed and studied. Observing that the original OVF from \cite{bando_dynamical_1995} results in frequent collisions in our initial simulations, we adopted the OVF studied by \citep{abdelhalim_real-time_2022} and added an additional safe distance $s_0$ factor to further prevent collisions for the final OVF given as

$$
    \tilde{V}(t) = v_{\text{opt}} \cdot \frac{\tanh(\frac{s(t) - s_0}{\theta} - \beta) + \tanh(\beta)}{1+\tanh{\beta}}
$$ \\
where \begin{itemize}
    \item $\beta$ is a tune-able parameter in the OVF
    \item $s_0$ is the minimum safe distance in $m$
    \item $v_{\text{opt}}$ is the desired velocity in $m/s$
    \item $\theta$ is a scaling factor in OVF
\end{itemize}

\subsubsection{FVDM}

Full Velocity Difference Model (FVDM) enhances the OVM model by additionally considering the relative velocity $\Delta s$. The FVDM uses identical OVF as OVM and predicts vehicle acceleration given as

$$
    a(t) = \alpha \left[ \tilde{V}(t) - v(t) \right] + \lambda \Delta s(t)
$$ \\
where $\lambda$ is the multiplier for relative speed factor \citep{jiang_full_2001} proposed as the primary improvement over OVM. 

\subsection{Data}

\subsubsection{Data Acquisition}
The Integrated Vehicle-Based Safety Systems (IVBSS) project, aimed at supporting the development safety features, results in a naturalistic driving database maintained at the University of Michigan Transportation Research Institute (UMTRI). This database includes naturalistic data recorded from 160 drivers over a two week period near Ann Arbor, United States. 


Each vehicle in this study is equipped with a radar mounted on the front bumper capable of measuring the instantaneous space headway to the lead vehicle at up to 150 meters. The vehicle's distance travelled since the start of the trip, as well as its current speed are obtained through the speedometer signal. An independent Inertial Measurement Unit (IMU) is used to measure the instantaneous longitudinal, lateral, and vertical accelerations. In addition, from the GPS data, the road type, categorized by highways, ramps, rural roads, and minor roads, is inferred.

For each identified congestion sequence, we query the fields as presented in Table \ref{tab:db-fields} to replicate lead vehicle behaviors in a virtual simulation environment. 

\begin{table}[h!]
\caption{Data Fields Queried And Computed}\label{tab:db-fields}
    \begin{center}
        \begin{tabular}{ l l l }
            \textbf{Name} & \textbf{Unit} & \textbf{Definition} \\ \hline
            Driver ID & - & Unique identifier for the driver  \\
            Trip ID & - & Unique identifier for the trip \\
            $t$ & s & Time elapsed since trip start \\
            $a(t)$ & $m/s^2$ & Ego longitudinal acceleration \\
            $v(t)$ & $m/s$ & Ego velocity \\
            $x(t)$ & $m$ & Ego distance traveled \\
            $s(t)$ & $m$ & Gap between ego and LV \\
            $\Delta s(t)$ & $m/s$ & Relative velocity between ego and LV \\
            Target ID & - & ID of LV, changes if LV changes \\
            AccelPedal & Percentage & Percentage of accelerator pedal applied \\
            BrakePedal & Binary & Whether brake pedal is applied or not
        \end{tabular}
    \end{center}
\end{table}

Upon querying congestion sequences using criteria mentioned in \ref{def-of-congestion}, we further split each congestion sequence into one or multiple Car-Following sequences: there may be multiple Car-Following events within a single congestion due to ego and lead vehicle lane changes or sensor malfunctioning, posing significant challenges in estimating lead vehicle trajectory. We further split each congestion sequence by the following criteria: 

\begin{enumerate}
    \item Driver ID and Trip ID is the same throughout each Car-Following event.
    \item Timestamps are continuous: $t_i - t_{i-1} = \Delta t = 0.1 \quad \forall i \in \mathbb{N}, 1 < i \leq n$
    \item Ego trajectory information ($x(t), v(t), a(t)$) is available for all $t$.
    \item Radar measurement is acquired ($s(t) > 0$). Since no crash occurred during the data collection, $s(t) = 0$ always indicates a measurement acquisition failure. 
    \item Target ID remains the same throughout. 
\end{enumerate}
This way, we ensure that each Car-Following sequence is a continuous time-series where the full trajectory of the same lead vehicle is available for simulation. 

\subsubsection{Data Processing}
Upon splitting the Car-Following events, we further refined data entries and computed additional quantities for simulation. Since the ego vehicle's distance travelled measurement is accurate up to 1 meter, the accuracy does not guarantee a smooth computation of the distance traveled by the lead vehicle, and thus the space headway during simulation. To address this issue, we numerically integrated the speed of the ego vehicle during each congestion period, then normalize it by the difference between the recorded distance measures to obtain a calibrated distance measure $\bar{x}(t)$. 

$$
    \bar{x}(t) = \frac{\bar{x}(n)-x(0)}{x(n)-x(0)} \left[x(0) + \int_{i=1}^n  v(i) dt \right]
$$ \\
We conduct numerical integration because we found the accuracy of the speed measurement to be more trustworthy since it is directly collected from wheel-speed sensor  with little to no slip. This allows us to obtain a normalized distance measure to compute a more smooth version of the ego, thus lead vehicle trajectory. 

Furthermore, we observed that the acceleration measured by the onboard IMU contains bias, most noticeable when the vehicle is stationary. However, we do not have additional data or capacity to apply advanced filtering to IMU measurements, so a basic bias correction for each Car-Following event is applied on the IMU readings. The bias in a unique congestion sequence is obtained as the average of IMU readings when the vehicle is stationary in that sequence (or zero, if vehicle is moving throughout). The bias is then subtracted from each IMU reading in that Car-Following sequence. 

$$
    \bar{a}(t) = a(t) - b_0 \quad \text{ where } b_0 = \frac{\sum_t a(t) \cdot [[v(t) = 0]]}{\sum_t [[v(t) = 0]]} \text{ is the average stationary drift}
$$ \\
To describe the trajectory of the lead vehicle, we computed the acceleration, velocity and distance travelled of the LV using information available in the IVBSS database. 

$$ 
    \begin{cases}
        v_\text{lv}(t) = v(t) + \Delta s(t) \\
        x_\text{lv}(t)= x(t) +s(t)
    \end{cases}
$$ \\
With the lead vehicle trajectory, we are then able to build a simulation framework using Newtonian kinematics. 

\subsection{Simulation}

The simulation of ego-vehicle behavior is designed and implemented using Newtonian kinematics, depending on each model's output. Since we do not have lateral information of the traffic or the driver's turning behavior, we assume that the drivers are driving in a perfectly straight line with no slope. While this is not true in all data recordings in the IVBSS database, we confirm that the road curvature and slope is negligible in all identified Car-Following sequences. The simulation is created using the following kinematic equations:

$$
    \begin{cases}
        a_P(t) = \text{ModelOutput}(*) \quad \text{ or } \quad \left[ v_P(t) - v_P(t-\Delta t)\right] / \Delta t \\
        v_P(t) = \text{ModelOutput}(*) \quad \text{ or } \quad v_P(t-\Delta t) + \Delta t \cdot a_P(t) \\
        x_P(t) = x_P(t-\Delta t) + v_P(t) \cdot \Delta t  + \left[a_P(t) \cdot (\Delta t)^2\right] / 2 \\
        s_P(t) = x_\text{lv}(t) - x_P(t) \\
        \Delta s_P(t) = \left[ s_P(t) - s_P(t-\Delta t) \right] / \Delta t
    \end{cases}
$$ \\
where $\Delta t$ is the discrete simulation time-step. All simulations are implemented with Python and PyTorch \cite{paszke_pytorch_2019}. It utilizes PyTorch's CUDA support to create tensor instances on both CPU and NVIDIA GPUs for fast parallel processing over different CF sequence and parameter sets. The simulation frequency is 10 Hz ($\Delta t = 0.1$), consistent with the frequency of data collection in the IVBSS project. The authors declare that graphing of simulation results is implemented with Matplotlib based on ChatGPT produced plotting code. 

\subsection{Model Fitting}

As Car-Following in the high-speed scenario has been extensively studied, we first employ parameters from existing literature that were calibrated with high-speed naturalistic datasets. We use these parameters in our simulator to observe how these parameters affect ego vehicle performance under low-speed Car-Following in congestions. 

Then, a calibration on the IVBSS congestion dataset is performed for each model to find the optimal parameters. When benchmarking and calibrating the models, we use the Root-Mean-Squared-Normalized-Error (RMSNE) criterion on the predicted space gap against the true space gap, as adopted by \citep{kesting_calibrating_2008}, \citep{zhang_comprehensive_2021} on Car-Following models. 

\label{eq:objective}
$$
    \text{RMSNE}(x_{\text{P}}(t), x_{\text{T}}(t)) = \sqrt{\frac{1}{n}\sum_1^n \left( \frac{x_{\text{{P}}}(i)-x_{\text{{T}}}(i)}{x_{\text{{T}}}(i)} \right)^2}
$$ \\
While other criteria are used for Car-Following calibration as well, such as RMSE in \cite{kim_identifying_2023}, \cite{vasconcelos_calibration_2014}, we choose RMSNE for its scaling that penalizes deviation more heavily at low ground truth (GT). This allows the calibrated models to perform very well at low-space gap (the majority situation in congestions). In exchange, we are more tolerant on the model output error at large ground truth space gap. 

Parameter calibration for Car-Following models involves finding the parameter set with the best simulation results. We employ the Genetic Algorithm (GA) to optimize the parameters for each model on our collective dataset. The genetic algorithm is an effective way of trying parameter combinations as explained by \citep{kesting_calibrating_2008} and adopted by field studies (\cite{vasconcelos_calibration_2014}, \cite{abbas_analysis_2011}). The Genetic Algorithm optimization is implemented using the PyGAD library \cite{gad_pygad_2024}. A total of 1024 parameters are evaluated over 400 mutation iterations for each model with a batch size of 256 for GPU-based parallel processing. 

\subsection{Clustering of Deviation Sequences} \label{sec:cluster-method}

Upon calibrating and analyzing simulation results, we further identify time segments where models and reality disagrees in space gap and space gap trend. Segments are identified manually using space gap and relative velocity RMSNE values as references. Segments where 1) the ground truth and models exhibit different trends and 2) the ground truth and models have large (single data point RMSNE > 1) differences in space gap predictions are selected to form the deviation dataset. Due to the requirement of time-series clustering, we label disagreement sub-sequences with at least 3-seconds of noticeable error. For sub-sequences with slightly smaller lengths, we select the time window by padding an equal amount of time on either side of the true erroneous sub-sequence. Sub-sequences less than one second is ignored due to the lack of interpret-able and cluster-able data. We select $m$ features from the dataset to form an $n$-sequence deviation dataset. 

After acquiring the deviation dataset, a truncated-sampling is performed to generate the necessary uniform dataset required by time-series clustering. The sampling process uniformly samples points in each selected sub-sequence at an interval of $t_i$ timestamps as starting-points of data chunks. Then, a chunk of length $k$ is sampled starting at each sub-sequence starting point. This returns sub-sequences that cover almost all of the deviation sequence and contains small overlaps to minimize the loss of information. 

Finally, a time-series clustering using Dynamic Time Warping (DTW) is performed on all sub-sequences. DTW is an algorithm to find the fitness between time-series data whose speed may differ. This is ideal in capturing commonalities in driver maneuvers which may differ in speed and response times. Our short time segments also allow DTW to work efficiently, even though it is generally considered computationally expensive. \citep{taylor_method_2015} uses DTW on driver behavior time-series to determine patterns in driver behaviors when they disagree with parametric models. Our DTW implementation is based on the work of \citep{tavenard_tslearn_2020}.

\section{Results}

\subsection{Data Processing}

In total, 160 unique congestion events from all drivers' trips, 24 of which are predominantly highway scenarios, which make up our dataset. Of the 24 congestions, 60 Unique Car-Following events are identified with continuous tracking of a lead vehicle, totaling 7440.8 seconds (2.07 hours) of recorded data. In preliminary testing, we observed that three out of the 60 Car-Following events contain large model-reality disagreement due to driver's anticipation of merge intent from vehicles in adjacent lanes. Since adjacent vehicle intent is not included in our dataset and not observed by any of the parametric models, we exclude these sequences from our final calibration. We do suggest, however, that Car-Following models consider driver's observations of adjacent vehicle maneuvers and intentions to generate more human-like responses. Besides, one other Car-Following sequence contains incorrect radar measurement due to the curvature of the road, therefore the lead vehicle trajectory could no longer be computed with fidelity. Various distraction factors, such as mobile phones, are also associated with large model-reality difference, but we chose to retain these sequences to observe how the parametric Car-Following models behave differently in these situations where drivers aren't actively responding differently due to other factors. A total of 56 sequences with 6975.1 seconds (1.94 hours), excluding the four mentioned above, are finally selected for model calibration and analysis. 

\subsection{Model Performance with Literature and Calibrated Parameters}
First, we report the simulation based on parameters found in existing literature. The parameter values calibrated in various literature for highway Car-Following can be found in Table \ref{tab:model-params}. 

\begin{table}[!ht]
\caption{Car-Following Model Parameters: Literature and Calibrated Values}
    \centering
    \begin{tabular}{l l l l l l l l }
        \hline
        Gipps & Parameter & $\tau$ (constant) & $v_{\text{opt}}$ & $a_{\text{max}}$ & $b_{\text{max}}$ & $s_0$ & $b^{\text{lv}}_{\text{max}}$ \\
        & Values by \cite{zhang_comprehensive_2021} & 1.02 & 41.88 & 1.24 & -2.57 & 7.83 & -2.00 \\
        & Calibrated Gipps & 1.02 & 12.48 & 3.49 & -10.00 & 1.77 & -9.43 \\
        
        \hline
        IDM, ACC & Parameter & $a_{\text{max}}$ & $b_{\text{comf}}$ & $s_0$ & $T$ & $v_{\text{opt}}$ & c (constant) \\
        & Values by \cite{zhang_comprehensive_2021} & 1.32 & 2.18 & 3.89 & 0.97 & 22.27 & 0.99 \\
        & Calibrated IDM & 1.06 & 0.50 & 4.30 & 2.05 & 40.00 & - \\
        & Calibrated ACC & 1.35 & 1.30 & 2.23 & 1.25 & 13.87 & 0.99 \\ 
        \hline 
        OVM, FVDM & Parameter & $\alpha$ & $\beta$ & $s_0$ & $v_{\text{opt}}$ & $\theta$ & $\lambda$ (FVDM) \\ 
        & Values by \cite{abdelhalim_real-time_2022},  & 0.195 & 0.10 & 4.00 & 36.13 & 9.41 & 0.20 \cite{wu_connections_2018} \\
        & Calibrated OVM & 0.83 & 0.39 & 2.83 & 15.18 & 12.24  & - \\
        & Calibrated FVDM & 1.02 & 0.003 & 2.29 & 22.04 & 29.70 & 0.001\\
        \hline
    \end{tabular}
    
    \label{tab:model-params}
\end{table}

\begin{figure}[!ht]
    \centering
    \begin{subfigure}[b]{0.48\textwidth}
        \centering
        \includegraphics[width=\textwidth]{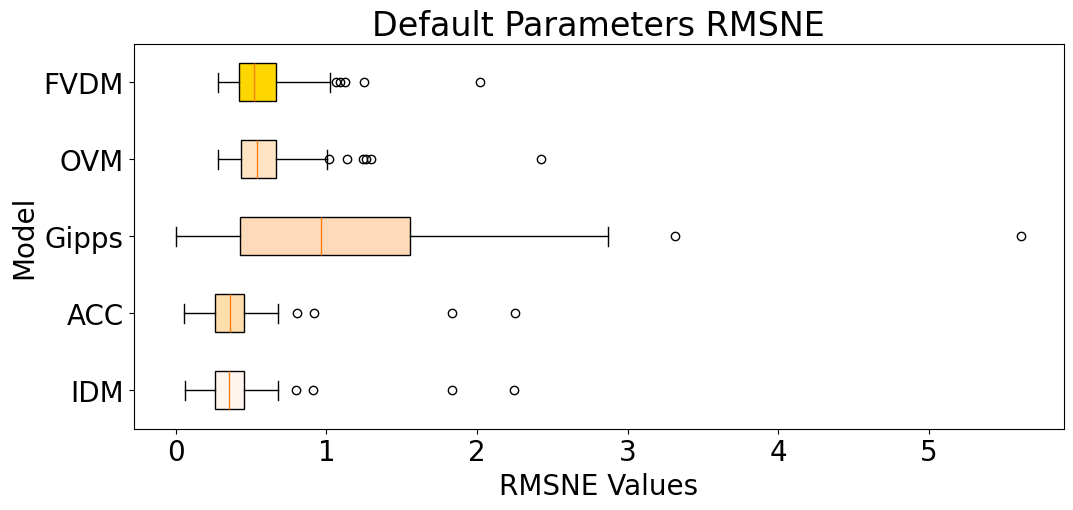}
        \caption{RMSNE distribution with Default Parameters}
        \label{fig:rmsne-def}
    \end{subfigure}
    \hfill
    \begin{subfigure}[b]{0.48\textwidth}
        \centering
        \includegraphics[width=\textwidth]{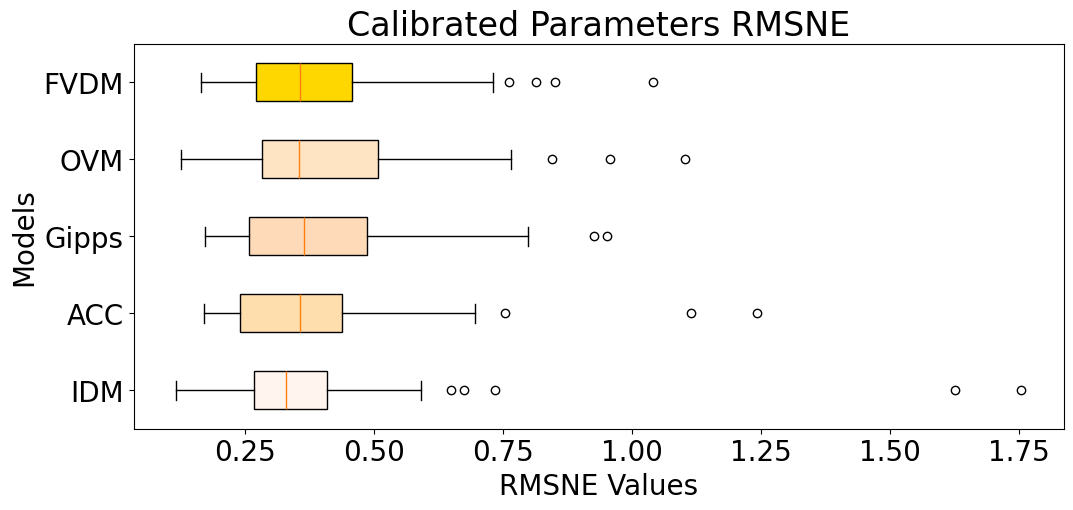}
        \caption{RMSNE distribution with Calibrated Parameters}
        \label{fig:rmsne-opt}
    \end{subfigure}
    \caption{Distributions of Per-Sequence RMSNE with high-speed and congestion optimized parameters. }
    \label{fig:param-rmsne-plot}
\end{figure}

\begin{figure}[!ht]
    \centering
    \begin{subfigure}[b]{0.48\textwidth}
        \centering
        \includegraphics[width=\textwidth]{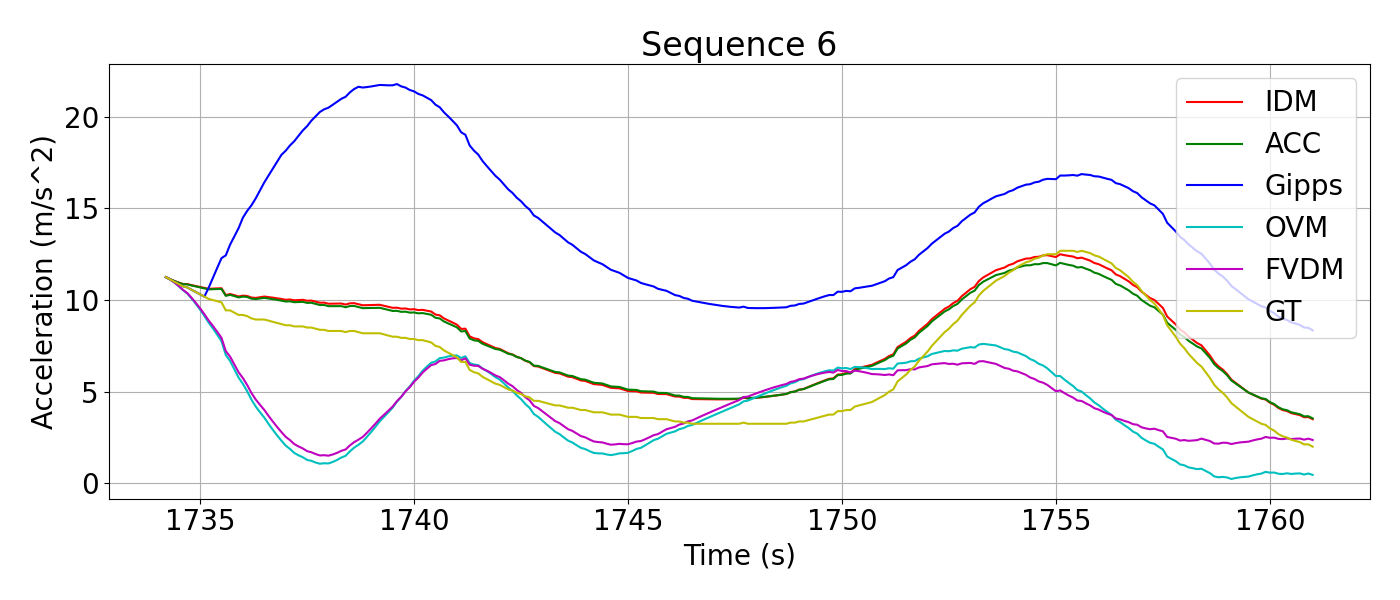}
        \caption{Space Gap of Default CF Sequence 6}
        \label{fig:def_s_6}
    \end{subfigure}
    \hfill
    \begin{subfigure}[b]{0.48\textwidth}
        \centering
        \includegraphics[width=\textwidth]{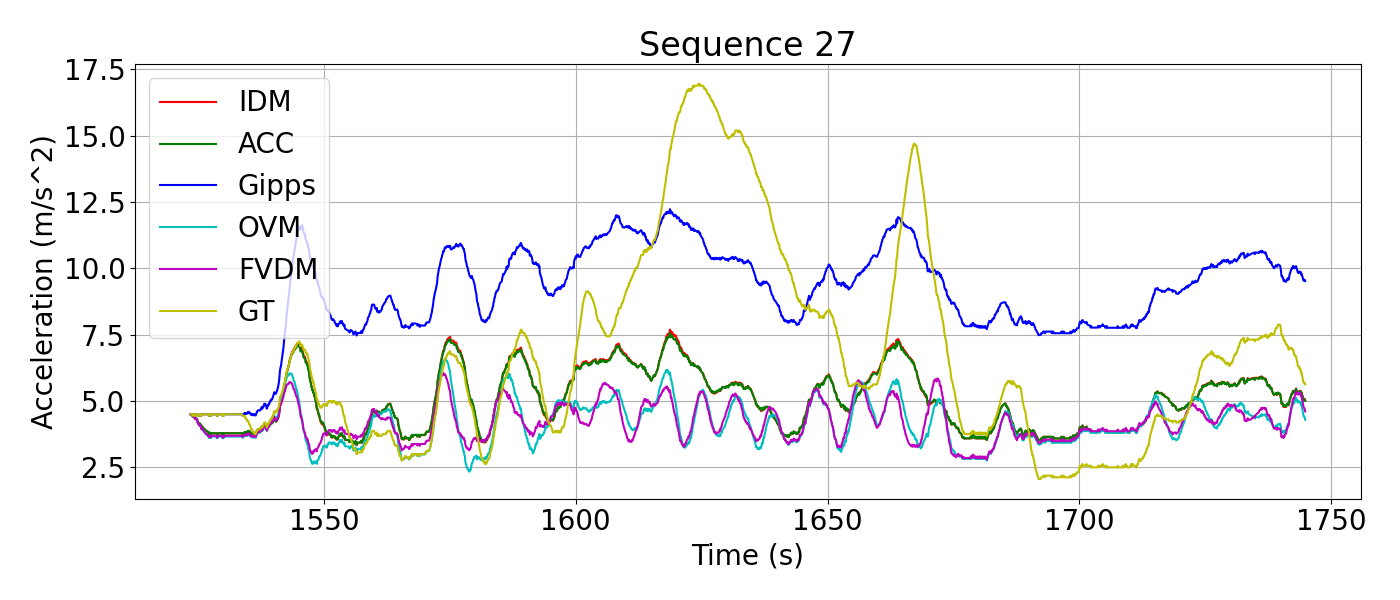}
        \caption{Space Gap of Default CF Sequence 27}
        \label{fig:def_s_27}
    \end{subfigure} 
    \\ 
    \begin{subfigure}[b]{0.48\textwidth}
        \centering
        \includegraphics[width=\textwidth]{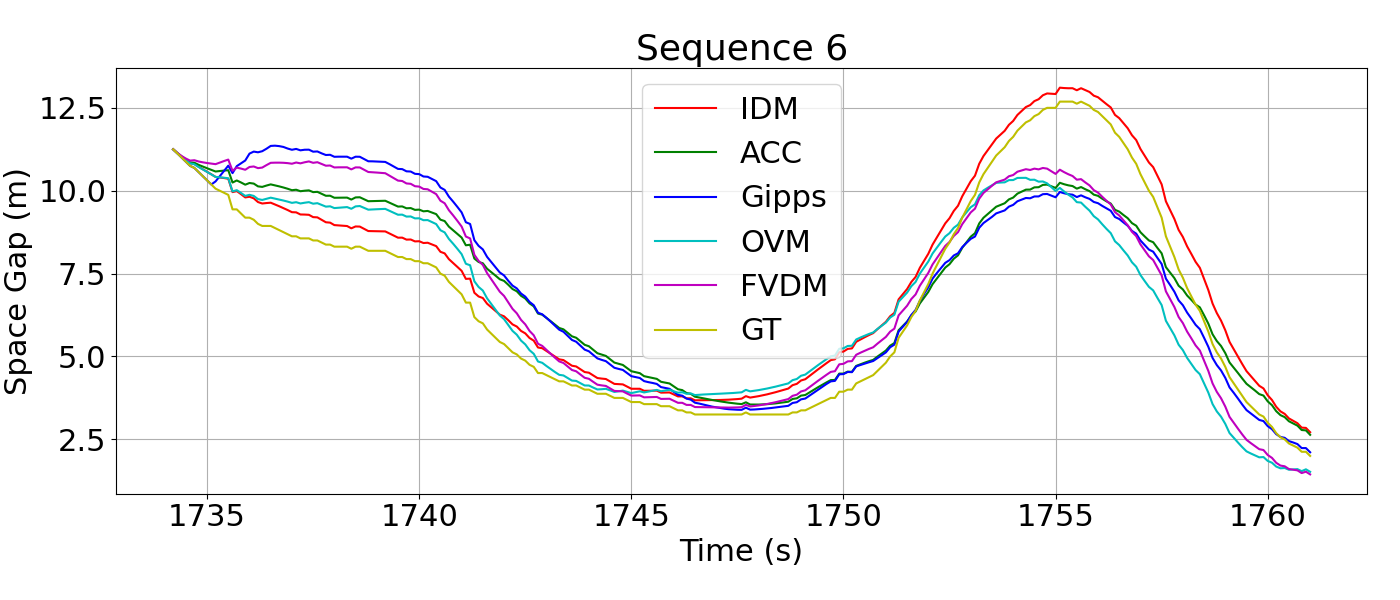}
        \caption{Space Gap of Calibrated CF Sequence 6}
        \label{fig:opt_s_6}
    \end{subfigure}
    \hfill
    \begin{subfigure}[b]{0.48\textwidth}
        \centering
        \includegraphics[width=\textwidth]{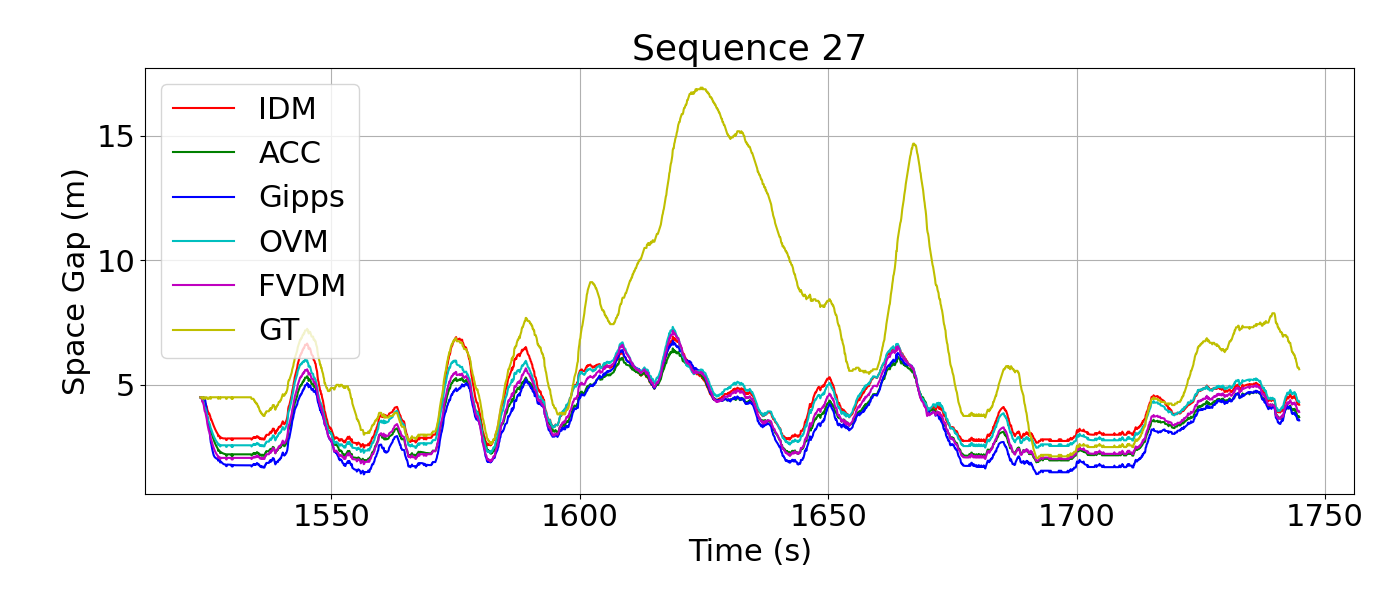}
        \caption{Space Gap of Calibrated CF Sequence 27}
        \label{fig:opt_s_27}
    \end{subfigure} 
    \caption{Example of Space Gap w.r.t Time before and after calibration on congestion dataset.}
    \label{fig:def-opt-example}
\end{figure}

Out of the five models, the parameters for Gipps model generated the largest RMSNE per sequence in general. This is primarily due to the minimum distance $s_0$ reported by \citep{zhang_comprehensive_2021} being too large for a congestion. As can be seen in Figure \ref{fig:def_s_27}, the gaps generated by Gipps for highway is unusually large compared to the ground truth. This also explains the large error mean and variance of Gipps RMSNE in Figure \ref{fig:rmsne-def}. The Gipps, IDM, and ACC models generally produce space gaps that resemble similar trends as the ground truth. Yet, OVM and FVDM responds slowly in acceleration output, causing a dampened-sinusoidal oscillation pattern in the predicted space gap, an example of which can be seen in Figure \ref{fig:def_s_6}.  

Upon calibration on the congestion dataset, all parametric models result in simulation results with an RMSNE error distribution that is smaller in both mean and variance, demonstrating the improved fit after calibration (Figure \ref{fig:rmsne-opt}). Examples of corresponding calibrated model output is shown in Figure \ref{fig:def-opt-example}. 

\subsection{Comparison between Models}

While the five models are designed based on three different heuristics, they exhibits very similar Car-Following gap patterns after calibration. The RMSNE error distributions (Figure \ref{fig:rmsne-opt}) confirm this observation, where all models has similarly distributed per-sequence RMSNE errors. Still, we notice the following differences in model behaviors: gap at standstill and possibility for collisions. 

\begin{figure}[!ht]
    \centering
    \begin{subfigure}[b]{0.48\textwidth}
        \centering
        \includegraphics[width=\textwidth]{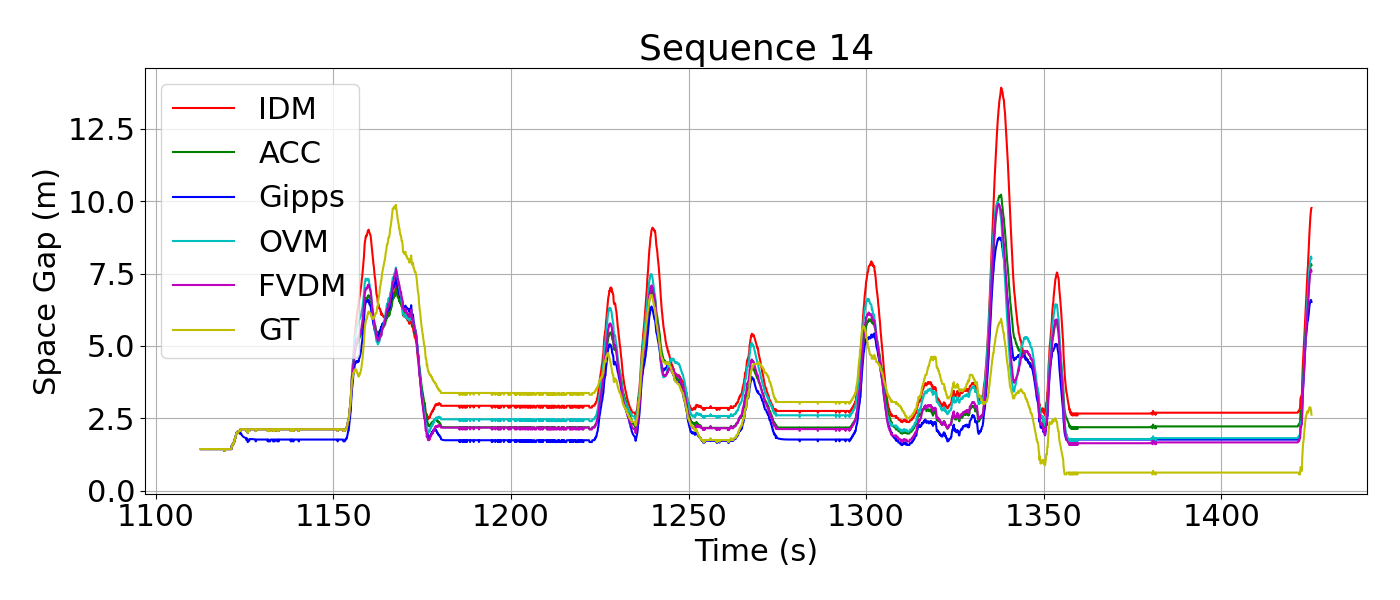}
        \caption{Space Gap of Calibrated CF Sequence 14}
        \label{fig:opt_s_14}
    \end{subfigure}
    \hfill
    \begin{subfigure}[b]{0.48\textwidth}
        \centering
        \includegraphics[width=\textwidth]{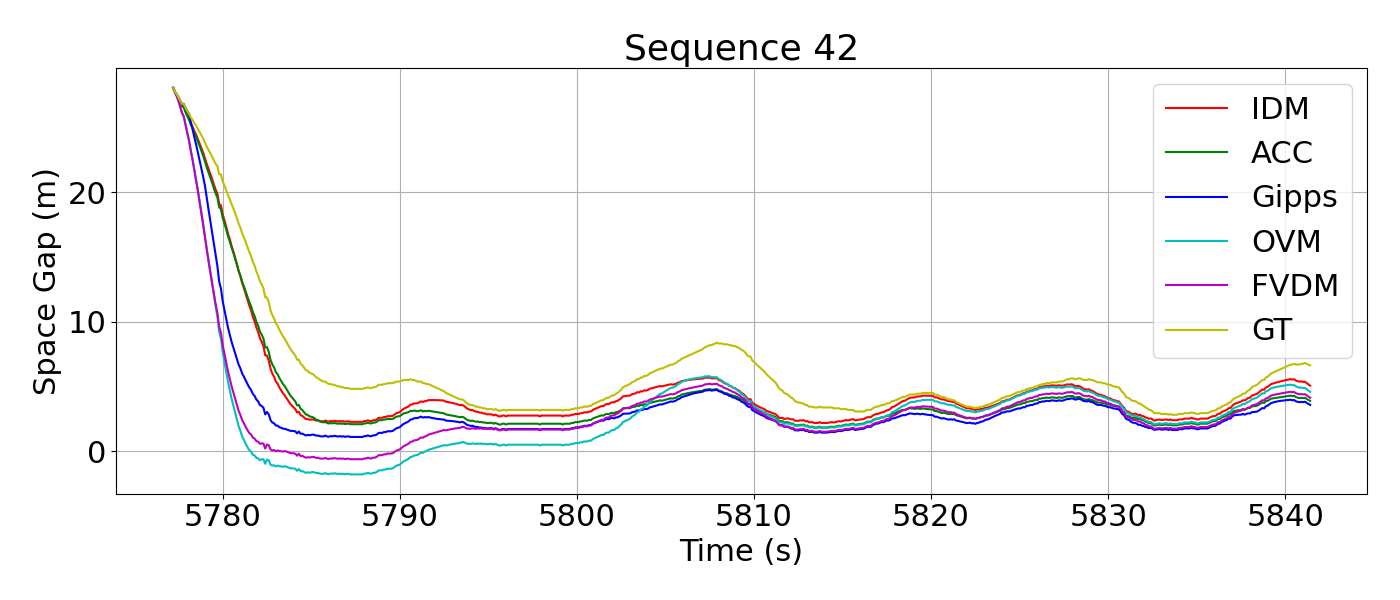}
        \caption{Space Gap of Calibrated CF Sequence 42}
        \label{fig:opt_s_42}
    \end{subfigure}
    \caption{Example Space Gap w.r.t Time for model-reality difference cases.}
    \label{fig:opt-mod-real-diff-collision}
\end{figure}

\subsubsection{Gap at Standstill}
Upon reaching zero velocity, all models follow a minimum safety gap, which generally agree with their respectively tuned $s_0$ value. Due to randomization and limited trials performed in model calibration, the safe following distance chosen by each model is slightly different. Still, since their values all fall within ranges of values documented in existing literature, we consider all of them reasonable and does not consider this a disagreement between models. We also notice that models are more consistent at staying close to the calibrated minimum distance gap upon reaching standstill, while human drivers are more flexible with their choices of stopping distance. The choice both between drivers and by the same driver, between trips or within the same trip can be different. Driver in Figure \ref{fig:opt_s_14} chose a stopping distance of 3.5 meters earlier in the congestion but chose to stop at just over 0.5 meters at the final segment, which can be uncomfortably small for other drivers. We suspect that the flexibility of human driver can be partially explained by the confidence and sense of control during coasting which will be discussed in \ref{sec:coasting}. 

\subsubsection{Collision}

Gipps, IDM, and ACC are all designed with an enforced minimium following distance to prevent collision. This factors is observed to be effective in our dataset, where these models produced zero collision with the lead vehicle. However, OVM and FVDM caused collision in two of the 60 CF sequences we simulated over. We suspect that the reason is a lack of hard minimum-distance constraint. Even after adding the safety distance factor in our OVF, it only produces a shift in the OVF output and produces a low target velocity. The constant multiplier $\alpha$ though, finally determines how much acceleration is applied, and the deceleration may not be enough in some cases to prevent a collision (Figure \ref{fig:opt_s_42}). 

Since the selected models perform very closely in cases other than those mentioned above, we perform the analyses in the next sections focusing on the difference between "model" and "reality", where "model" represents the general behavior of all models. 

\subsection{Model-Reality Comparison}

Although the five parametric models largely agree with each other, 
Further review of video and telemetry data in sequences where model and reality disagree reveal the following driving patterns by real-world drivers: 

\subsubsection{Driving with Momentum}

\begin{figure}[!ht]
    \centering
    \begin{subfigure}[b]{0.48\textwidth}
        \centering
        \includegraphics[width=\textwidth]{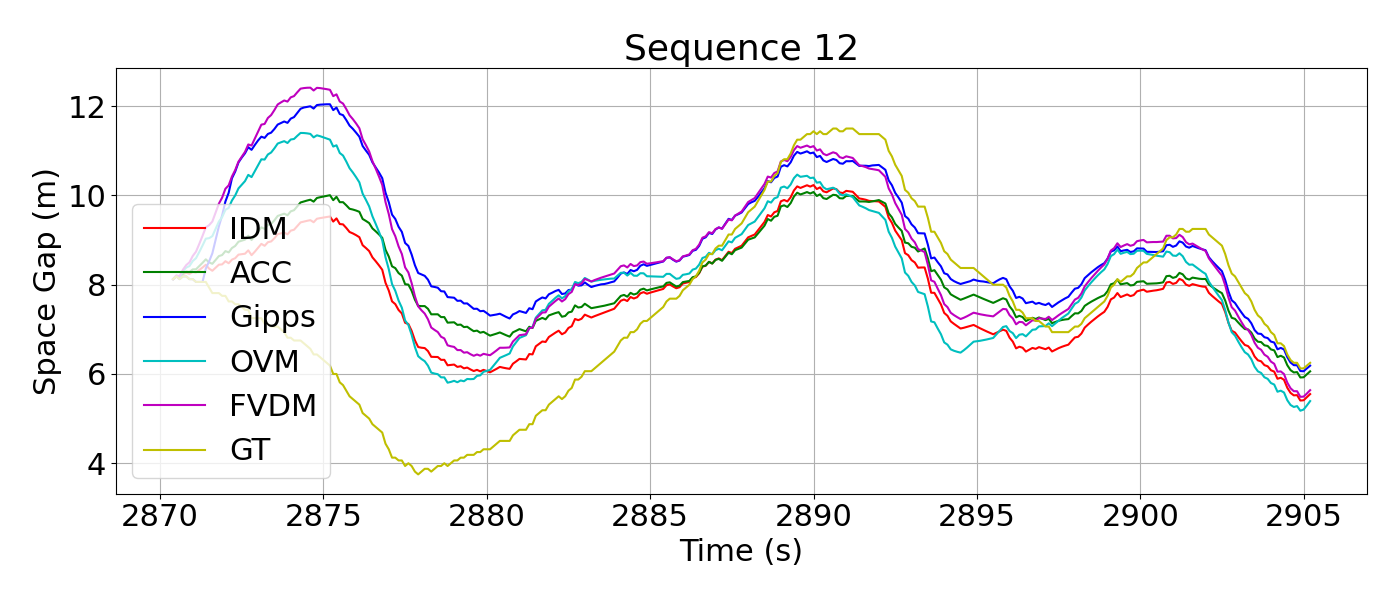}
        \caption{Space Gap of Calibrated CF Sequence 12}
        \label{fig:opt_s_12}
    \end{subfigure}
    \hfill
    \begin{subfigure}[b]{0.48\textwidth}
        \centering
        \includegraphics[width=\textwidth]{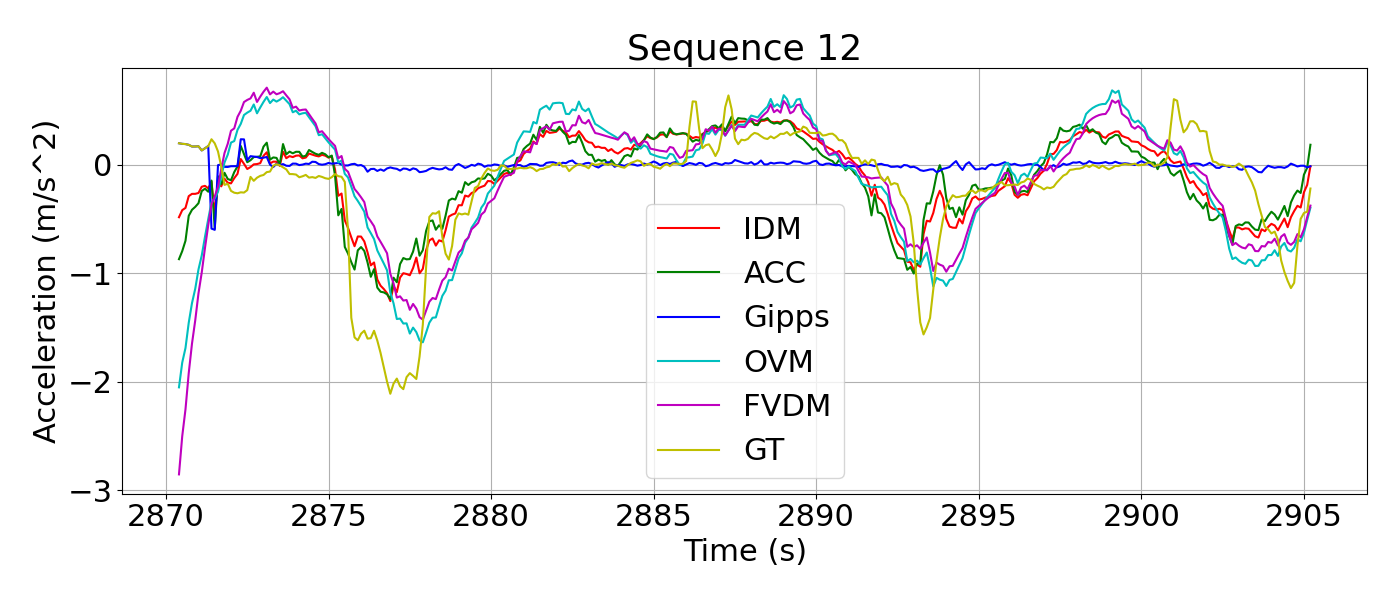}
        \caption{Acceleration of Calibrated CF Sequence 12}
        \label{fig:opt_a_12}
    \end{subfigure} 
    \caption{Example Space Gap/Velocity w.r.t Time for model-reality difference cases. The first subfigure shows the space gap case where driver drives with momentum, and the second subfigure illustrates the acceleration pattern adopted by the driver and models. }
    \label{fig:opt-mod-real-diff-momentum}
\end{figure}

All five Car-Following models are designed with some minimum safe-distance in mind. Thus, if a Car-Following sequence starts with a space gap smaller than parametric models prefer, they actively control the vehicle to slow down and maintain a safe distance (as seen in opening seconds of Figure \ref{fig:opt_a_12}). Human drivers, on the other hand, tend to keep the velocity at entry and rarely brake to actively maintain a safety distance. This momentum, however, can catch the driver unprepared and was forced to brake harder. In Figure \ref{fig:opt_s_12}, for example, after an ego lane change, the driver did not immediately slow down as the models did to maintain a safe distance. When the lead vehicle braked immediately after, the space gap quickly reduced, forcing the driver to rapidly decelerate around 2075 seconds in Figure \ref{fig:opt_a_12}. In comparison, the acceleration curves produced by the parametric models are much smoother. 

\subsubsection{Coasting} \label{sec:coasting}
\begin{figure}[!ht]
    \centering
    \begin{subfigure}[b]{0.48\textwidth}
        \centering
        \includegraphics[width=\textwidth]{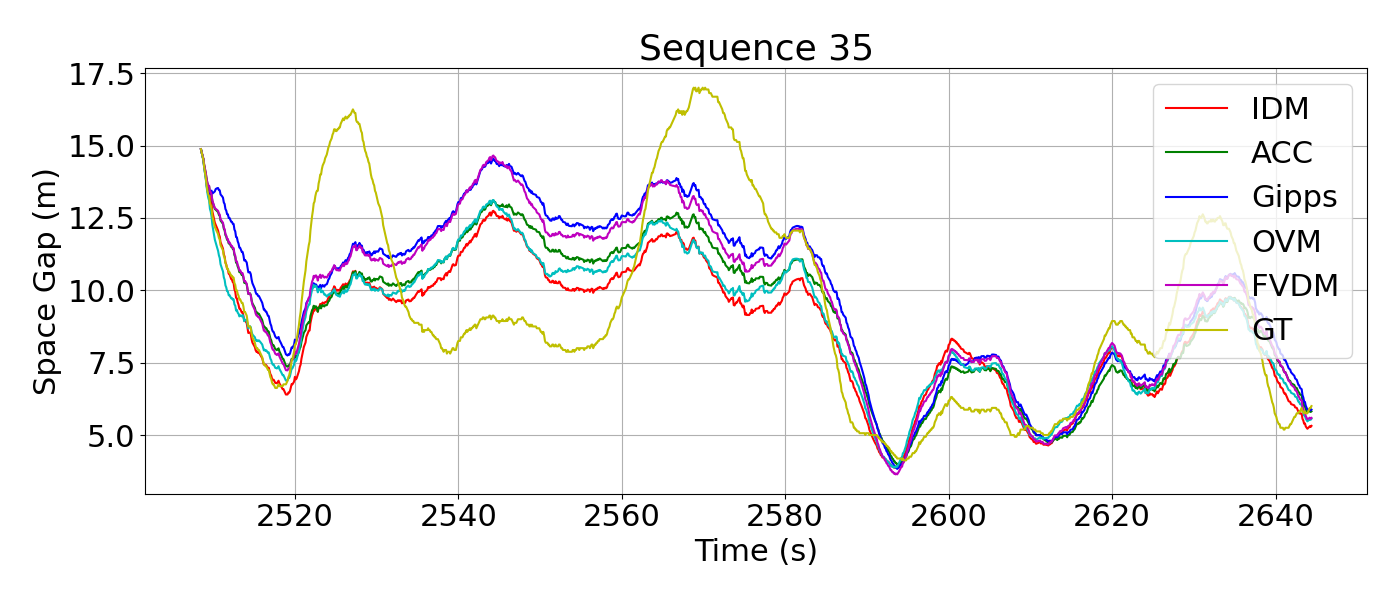}
        \caption{Space Gap of Calibrated CF Sequence 35}
        \label{fig:opt_s_35}
    \end{subfigure}
    \hfill
    \begin{subfigure}[b]{0.48\textwidth}
        \centering
        \includegraphics[width=\textwidth]{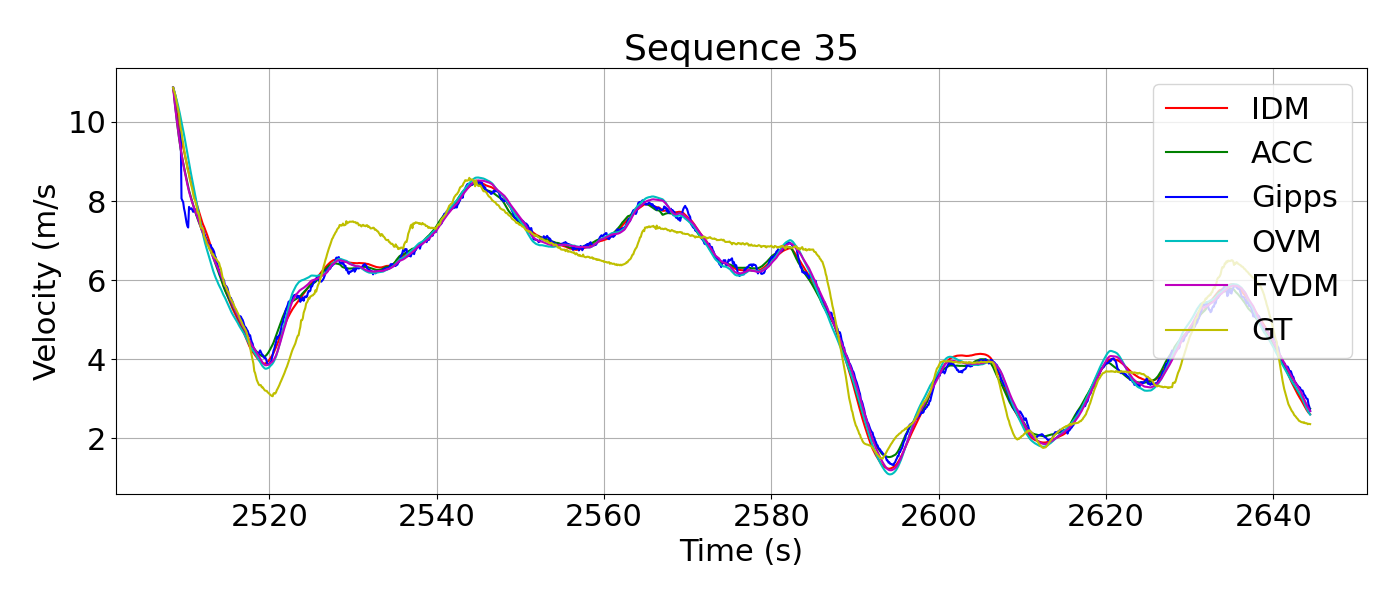}
        \caption{Velocity of Calibrated CF Sequence 35}
        \label{fig:opt_v_35}
    \end{subfigure} 
    \caption{Example Space Gap/Velocity w.r.t Time for model-reality difference cases. Driver in this sequence elects to lift and coast as visible in subfigure 2. }
    \label{fig:opt-mod-real-diff-coast}
\end{figure}

By reviewing video telemetry data, we notice that drivers often employ the "lift and coast" strategy during a congestion, where driver applies neither pedals at high speed to use the vehicle's momentum to move forward. A vehicle with internal-combustion engine in this case has only a minor deceleration from transmission friction, causing a minor near-constant deceleration, which the driver can easily predict. Parametric models, on the other hand, produce more smooth acceleration values and frequently switches between acceleration and deceleration with no delay. As seen in Figure \ref{fig:opt_s_35}, the large gap between time 2560s and 2580s is caused by the driver's lift-and-coasting starting at 2550s in Figure \ref{fig:opt_v_35}. The brief increase in driver's speed indicates that the driver applied acceleration briefly, and then resumed costing. During this period, all of the models actively track the lead-vehicle and output speed curves that fails to match the ground truth. 

\subsubsection{Idle Creep} \label{sec:idle-creep}
\begin{figure}[!ht]
    \centering
    \begin{subfigure}[b]{0.48\textwidth}
        \centering
        \includegraphics[width=\textwidth]{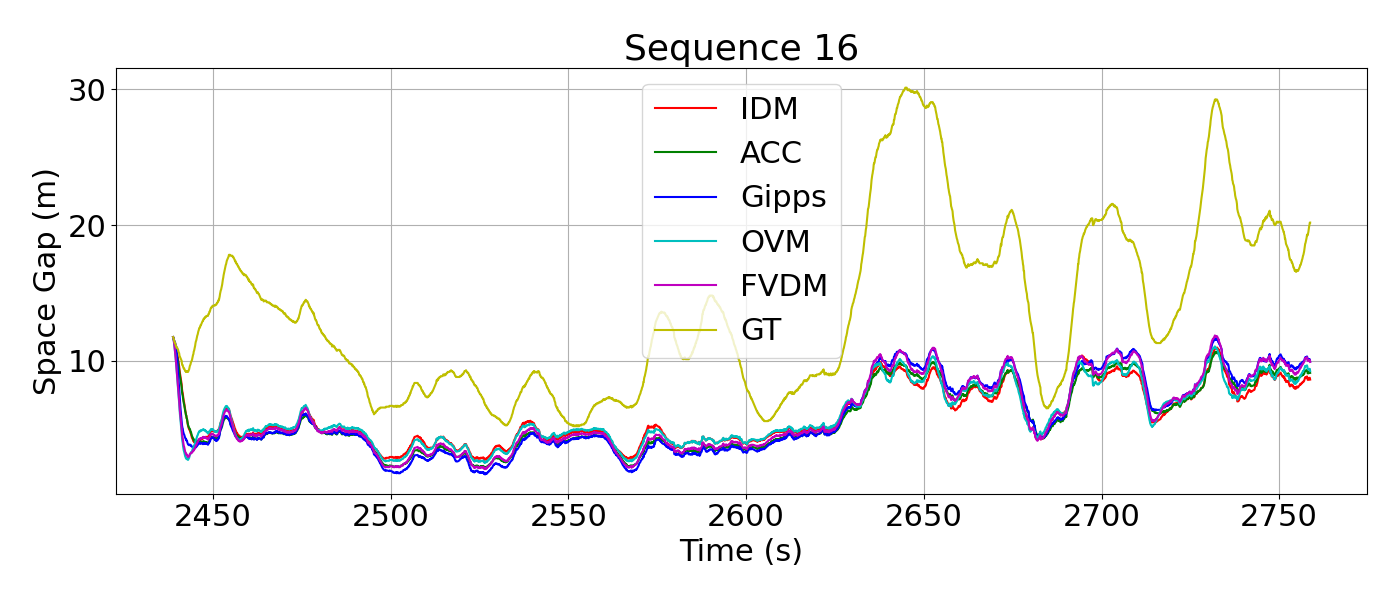}
        \caption{Space Gap of Calibrated CF Sequence 16}
        \label{fig:opt_s_16}
    \end{subfigure}
    \hfill
    \begin{subfigure}[b]{0.48\textwidth}
        \centering
        \includegraphics[width=\textwidth]{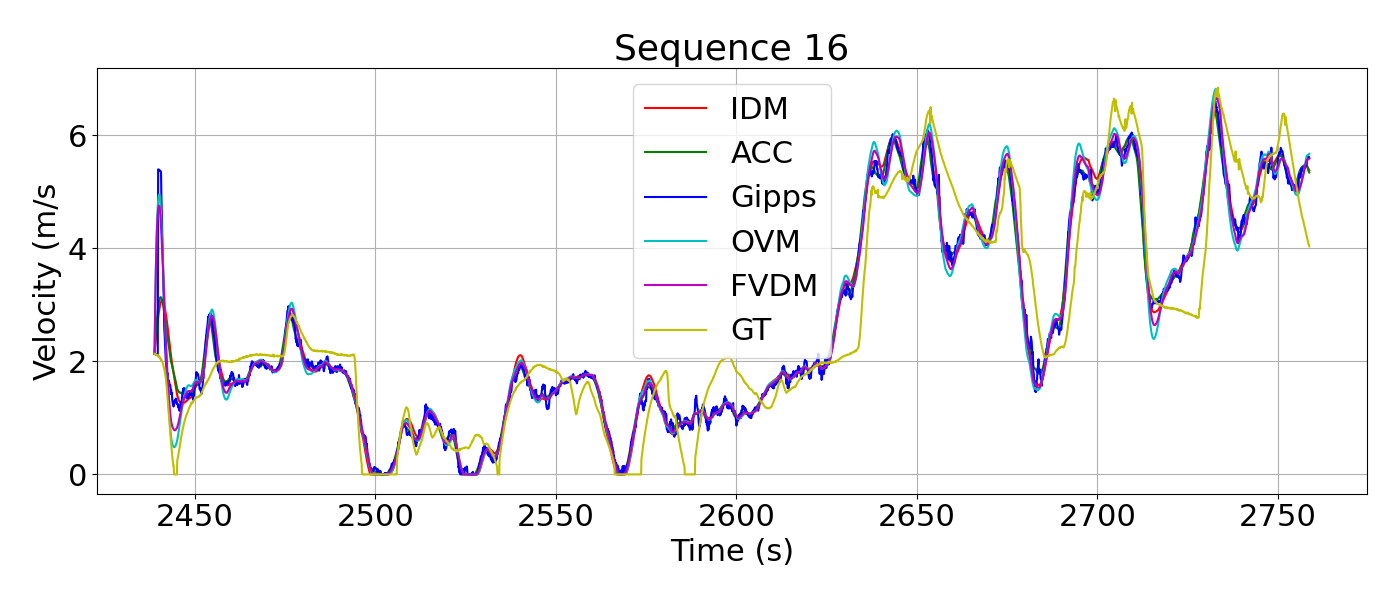}
        \caption{Velocity of Calibrated CF Sequence 16}
        \label{fig:opt_v_16}
    \end{subfigure} 
    \caption{Example Space Gap/Velocity w.r.t Time for model-reality difference cases. Driver in this sequence elects to use idle creep as visible in subfigure 2. }
    \label{fig:opt-mod-real-diff-idle}
\end{figure}

Another unique property of internal-combustion engine vehicles that drivers take advantage of in low speed Car-Following is the "idle creep", where the vehicle spontaneously provides power to drive vehicle to around 1.5m/s (our dataset shows a maximum of 1.39 m/s before driver pressed accelerator pedal on a 2007 Honda Accord ) without driver input in the accelerator pedal. Drivers often use this feature to slowly creep back to the back of the lead vehicle in slow congestion scenarios with their foot consistently on the brake pedal. As seen in Figure \ref{fig:opt_s_16}, the driver rarely ever spontaneously accelerate between 2550 and 2700s. Instead, the idle power is used to driver the car forward at creeping speeds. This allows the driver to be free from constant pedal application in congestion when they value space gap less. This particular driver take advantage of idle creeping in multiple Car-Following sequences in our dataset, but due to the limited sample size, we are unable to draw any conclusion as to which types of drivers are more likely to use idle creep in a congestion, and in what situations. 

\subsection{Clustering for Disagreement Sections}

The time-series clustering analysis we report below is based on DTW results with length $k = 30, t_i = 20$ and $m = 3$ features: acceleration pedal percentage, brake pedal application (binary), and calibrated acceleration. A total of 1277 sub-sequences are created from the labelled ranges. Attempts to cluster sequences with shorter lengths ($k = 15, t_i = 10$, and $k = 10, t_i = 7$, respectively) as well as clustering more features (with the addition of speed, relative speed, space gap as features), and the combination of length 30 sequence with three features produce the most interpret-able clusters.  We find $k = 30$ timestamps, or 3 seconds of data, a healthy amount to describe a short-time trend in the driver's behavior or change in situation. A shorter sub-sequence may fail to capture short-term changes and only capture an instantaneous snapshot of the driver status. The reason not to cluster speed, relative speed, or other factors is that we intend the clustering to reveal patterns of human activity in order to then reverse-engineer the conditions in which human drivers respond in those patterns, rather than clustering the conditions in the first place. We found, in practice, that including environment variables, such as speed and space gap, dilutes the recognition of human choices, as each dimension is weighted equally in multi-dimension DTW. 

\begin{figure}[!ht]
    \centering
    \begin{subfigure}[b]{0.3\textwidth}
        \centering
        \includegraphics[width=\textwidth]{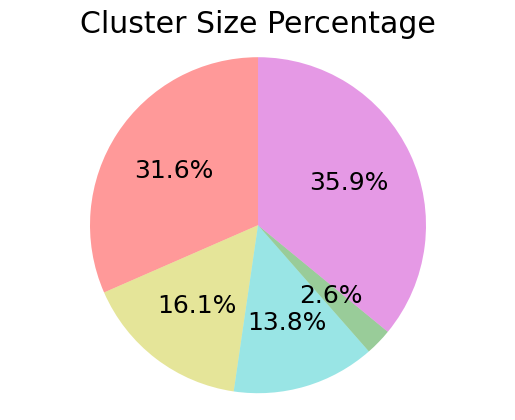}
    \end{subfigure}
    \hfill
    \begin{subfigure}[b]{0.67\textwidth}
        \centering
        \includegraphics[width=\textwidth]{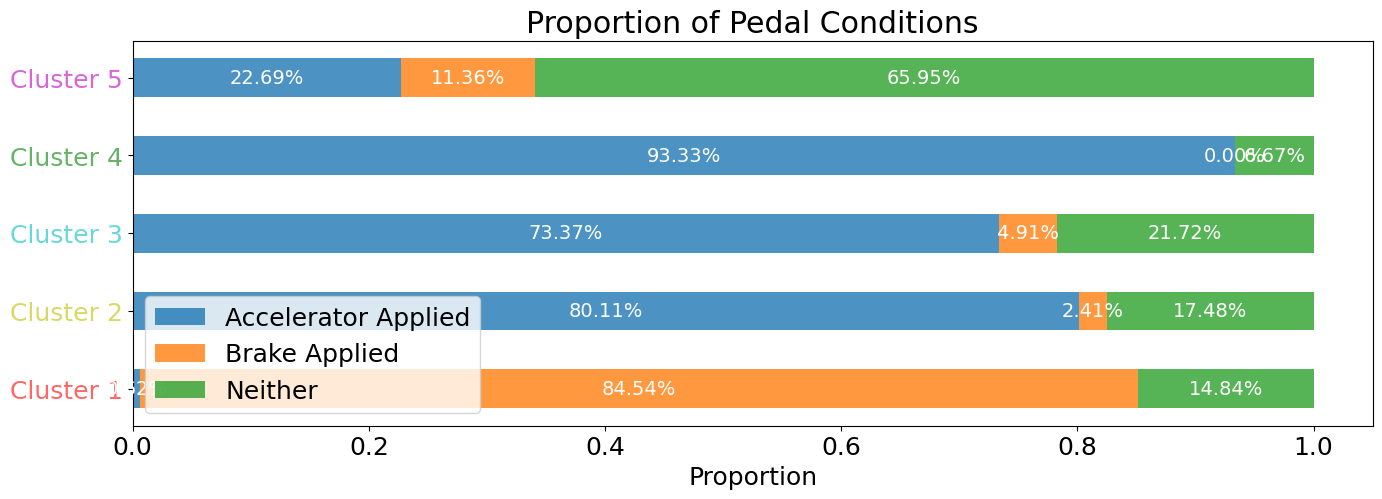}
    \end{subfigure}
    \caption{Results of time-series clustering. The first subfigure shows the size percentage of each cluster, and the second subfigure shows the pedal application distribution for each cluster. }
    \label{fig:cluster_res_1}
\end{figure} 

Using the "elbow method" on length 30 sub-sequences with three features, five clusters is determined to be an optimal amount of clusters for our selected dataset. Figure \ref{fig:cluster_res_1} shows the distribution of clusters and their pedal application patterns found in the time-series data. Clusters 2,3,4 primarily focus on acceleration pedal application. Cluster 1 is primarily identified as a braking cluster. Notice that since our brake pedal data is binary (either applied or not), we only obtained one cluster for braking behavior whereas this may not be the case had we been offered brake percentage data as well. Future research could provide more details into the braking cases should the data be available. Cluster 5 represents the case where the driver is not applying either pedals most of the time. This is when driver is either lift-and-coasting or idle-creeping as discussed above, depending on the speed range. 

\begin{figure}[!ht]
    \centering
    \begin{subfigure}[b]{0.99\linewidth}
        \centering
        \includegraphics[width=1\linewidth]{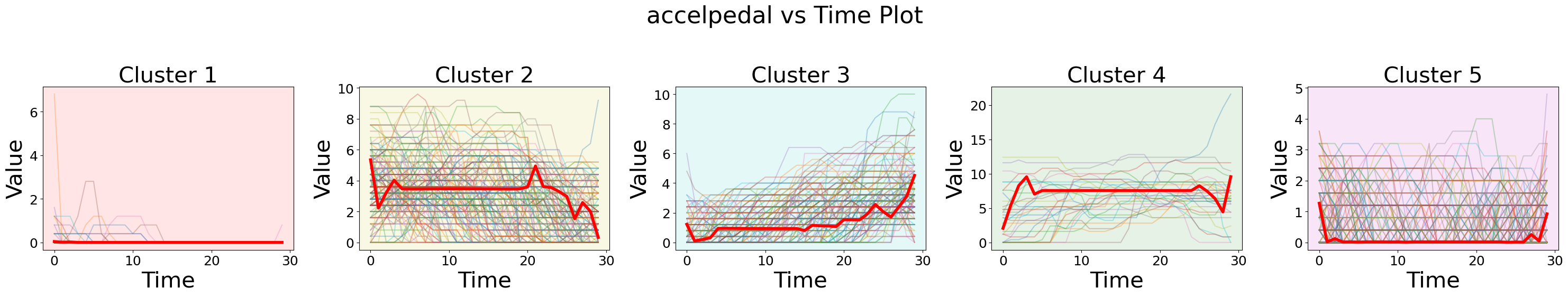}
        \caption{Accelerator Pedal Clustering Results and Centroids}
        \label{fig:cluster-accel-ts}
    \end{subfigure} 
    \\
    \begin{subfigure}[b]{0.99\linewidth}
        \centering
        \includegraphics[width=1\linewidth]{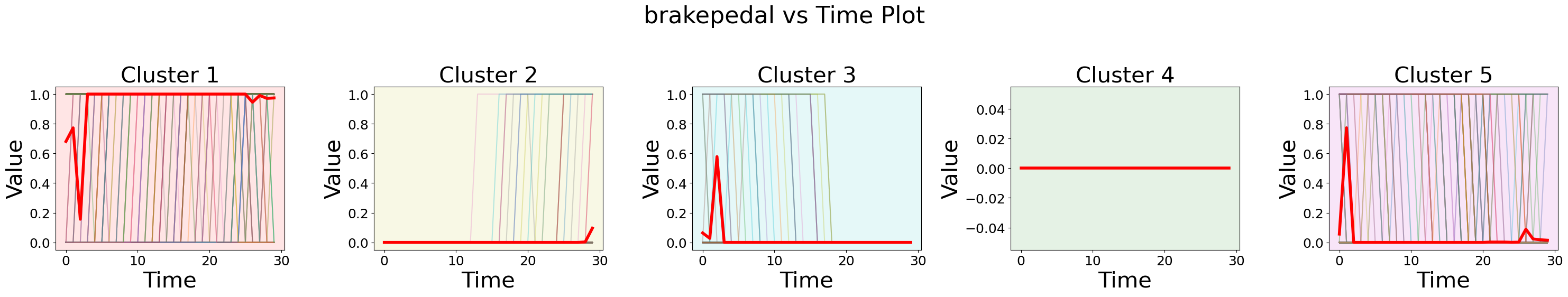}
        \caption{Brake Pedal Clustering Results and Centroids}
        \label{fig:cluster-brake-ts}
    \end{subfigure} 
    \caption{Resulting time-series stumps of clustering. Each of the clustered sub-sequence stumps are plotted along with the bolded red centroids.}
    \label{fig:cluster-ts}
\end{figure}

Figure \ref{fig:cluster-ts} shows the accelerator and brake pedal applications upon clustering and their respective centroids. Cluster 1 centered around full brake pedal application and Cluster 5 centers around no application of either pedals, confirming previous observations. While cluster 2, 3, 4 have similar percentage of accelerator pedal usage, Figures \ref{fig:cluster-accel-ts} and \ref{fig:cluster-brake-ts} reveals the different trends they capture, Cluster 2 is characterised with a decrease in accelerator pedal application and increase in braking, representing a transitional phase from accelerating to braking. Cluster 3 is the opposite of Cluster 2, representing a transition from brake pedal application to acceleration. Finally, Cluster 4 contains no braking record, representing full acceleration phases of human driving. 

\begin{figure}[!ht]
    \centering
    \includegraphics[width=0.7\linewidth]{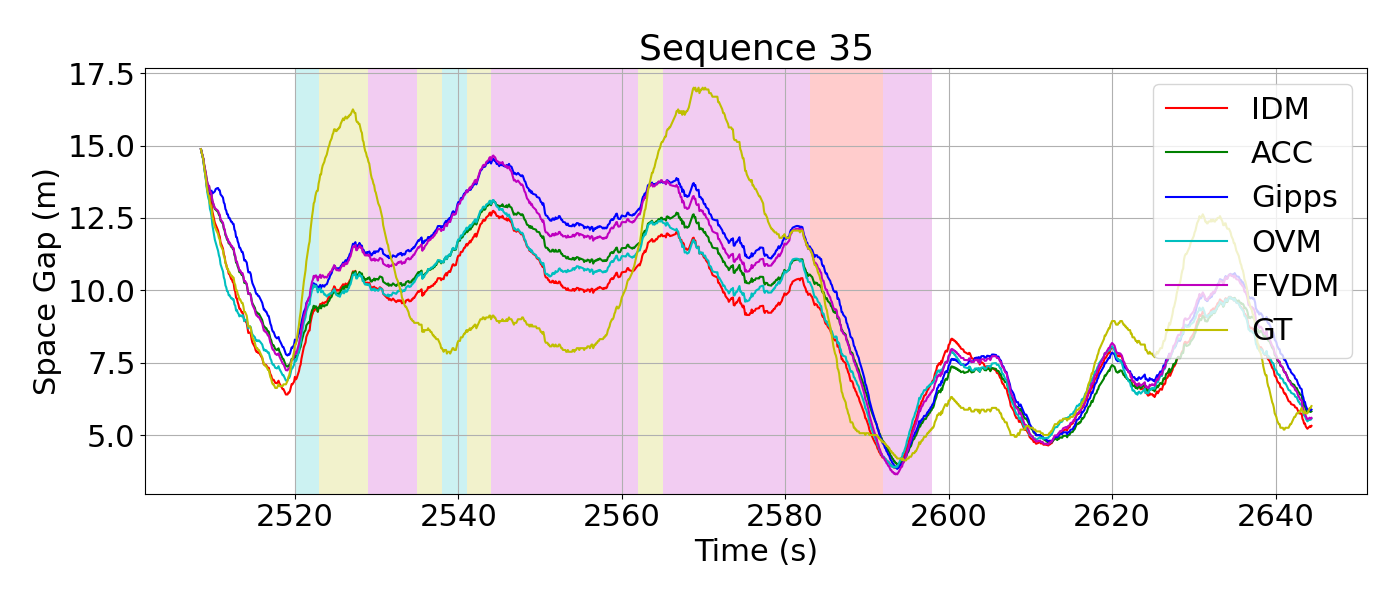}
    \caption{Space gap of Calibrated Sequence 35 with clustering results overlapped. Each of the color strips represent a clustering results of that time-series chunk with the colors matching definitions in Figure \ref{fig:cluster_res_1}}
    \label{fig:opt_s_35_color}
\end{figure} 
\begin{figure}[!ht]
    \centering
    \includegraphics[width=0.7\linewidth]{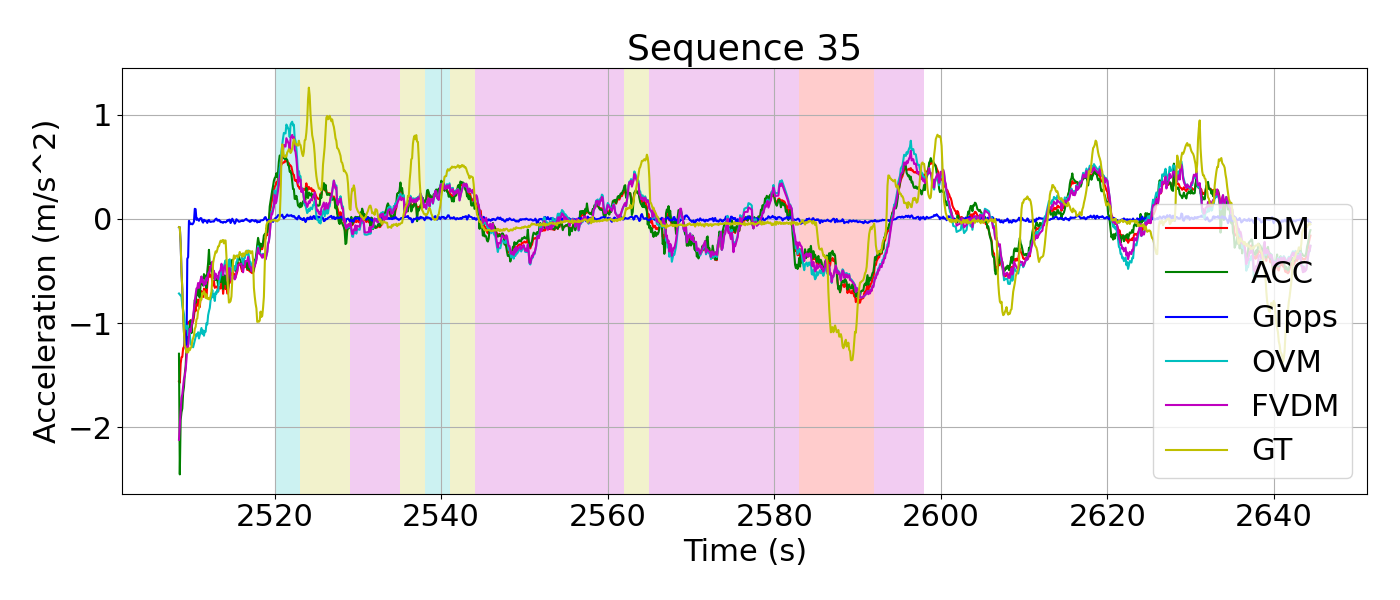}
    \caption{Acceleration of Calibrated Sequence 35 with clustering results overlapped. Each of the color strips represent a clustering results of that time-series chunk with the colors matching definitions in Figure \ref{fig:cluster_res_1}}
    \label{fig:opt_a_35_color}
\end{figure}

Figure \ref{fig:opt_s_35_color} shows an example of overlap of model-reality space gap plot and segment clustering results. In this particular sequence where none of the parametric models were able to accurately predict the trajectory of the ego vehicle, the driver drives with Cluster 5 style (coasting) in over half of the time. Among extended lift-and-coasting, this driver conducts short bursts of acceleration or braking, visually distinct between 2540 and 2580s in Figure \ref{fig:opt_a_35_color}, rather than a continuous acceleration curve produced by the parametric models. Other Car-Following sequences exhibit similar trends with a large percentage of Cluster 5 or Cluster 1 along with intermittent accelerations. 

Overall, evidence suggests that the five parametric models, when calibrated using the same metric on our congestion dataset, performs close in predicted space gap to lead vehicle. Nevertheless, they unanimously fail in certain scenarios where drivers frequently exhibit coasting and idle creeping behaviors as suggested by both IVBSS video data and driver maneuver clustering. 

\section{Discussion}

\subsubsection{Summary}

This paper thus far investigated the performance of five parametric Car-Following models under low speed congestions. Model performance is found to be comparable and similar after calibration. Then, series of driver behaviors that disagree with model output is identified and investigated. Several factors are empirically observed to be associated with noticeable model-reality gap. Sequences are further clustered using Dynamic Time Warping to form five clusters of driver maneuvers commonly conducted when model-reality gap is present. Finally, the meaning and implications of these clusters are discussed. 

\subsubsection{Model Behaviors under congestion}

Even though the Gipps model is proposed decades earlier than the other four models used in this study, its robustness and accuracy is comparable to the other four models after calibration. The performance comparison between Gipps and IDM in our simulation results agree with those found by \citep{zhang_comprehensive_2021} using a mixed road-condition dataset. Our RMSNE values from simulations with calibrated models also align with those reported in \cite{zhang_comprehensive_2021}, indicating a consistency with these Car-Following models applied to different datasets. 

While \citep{jiang_full_2001} showed that the addition of relative speed improved the accuracy and eliminated unrealistic accelerations from OVM, this factor poses a negligible improvement in our calibrated simulations compared to OVM. We notice that the calibration results in a factor $\lambda = 0.001$, the minimum of the search range values. This indicates that the compact distribution of relative speed in congested scenarios diminishes the impact of this factor in FVDM, making it virtually identical to OVM in all simulations. We suspect that this factor may be more useful in situations with significantly larger relative velocities, such as high speed initial approaches to a congestion, but those scenarios are extremely limited in our dataset. Without special attention, the general calibration process will not specifically optimize for such scenarios. We suggest that separate calibration for the $\lambda$ parameter be conducted on specific sub-datasets, should FVDM be used in congestion behavior simulations. 

\subsubsection{Calibration Criterion}

The Calibration criterion (RMSNE) selected for this study may have contributed to the converging behaviors of the models. Choosing a normalized metric over the space gap may have over-penalized model-reality gaps in the low gap sections. Whereas in high speed Car-Following, the minimum space gap to lead vehicle is relatively large, this distance may become extremely small in certain congestion events (Figure \ref{fig:opt_s_14}), making the ratio between largest and smallest space gap potentially more extreme than in free flow Car-Following. This penalty could have cause the parameter selection to converge on optimizing low-space gap behaviors, rather than providing the freedom for each model to excel in their respective ways. Still, we think the theoretical discussion provided by \citep{kesting_calibrating_2008} applies in congestions, and  we found no clear theoretical or empirical superiority is found with other metrics like RMSE or MSE. Future work into objective selection or multi-objective optimization for congestion Car-Following may result in more distinct model behaviors and characteristics. 

\subsubsection{Driver Behavior under congestion}

Coasting and idle creep are common features of internal-combustion-engine vehicles that drivers take advantage of. Yet, neither have been extensively studied in parametric Car-Following models.

\citep{kim_asymmetric_2024} recently proposed the ARM model that integrates human's "coasting behavior" as a transition phase between acceleration and deceleration. Authors concluded that the ARM model clearly outperformed baseline models including IDM and Gipps, which are also evaluated in this paper. This shows the efficacy of including coasting behavior as an important human factor in Car-Following models. Nevertheless, \citep{kim_asymmetric_2024} emphasized their model's regime design on high speed scenarios and evaluated their model on the NGSIM dataset, primarily composed of highway recordings. Whether the ARM model integrates coasting the way we find drivers do in congestions requires further investigation. 

Furthermore, idle creeping has largely been ignored by Car-Following models for its extremely limited usage in low speed. This is, however, a preferred way for some drivers to conduct Car-Following in low-speed congestions for its effortless control: driver needs not applying either pedal for the car to creep back to the lead vehicle in a controlled manner. \cite{timbario_testing_2022} and \cite{timbario_engine_2023} investigated the calibration and proposed a simulation method for idle creeping behavior on a variety of vehicles. They suggest that while idle creep measurement is consistent through multiple runs and multiple vehicles of the same make, but authors fail to correlate idle creep with another known vehicle parameter. Additionally, idle creep is also different from vehicle to vehicle. This may require a parametric model including such parameter further calibration for each make of vehicle, which is not always feasible. 

\subsubsection{Limitations}

This study is limited by the amount and diversity of data available in the IVBSS project, such that we are unable to reliably obtain extended Car-Following sequences in city-traffic as we are in highway environments. Thus, the observations and calibration results presented in this study are limited to highway congestions and may not apply to urban-traffic situations. The IVBSS dataset from 2011 also contains sensor readings with limited accuracy (particularly the raw IMU reading) and limited sensor capabilities (particularly the binary brake signal acquisition) such that the accuracy and generalizability of the results is impacted. To build a viable simulation environment, extra data processing and estimation including bias removal and numerical integration was conducted, which may unavoidably distort data fidelity. Should more comprehensive and accurate datasets be available, future research could be conducted to reevaluate our findings.

This study is also limited to five kinematics-based Car-Following models, and only model-reality differences where these kinematic models fail are discussed. While no psychology or control theory based models, to the best of our knowledge, explicitly models human drivers' coasting and idle creeping behaviors, we could not conclusively state that those models would fail under these circumstances as well. Nevertheless, coasting and idle creeping are kinematic features of internal combustion engine vehicles and have the potential to be efficiently integrated into the existing parametric models' frameworks to improve their accuracies. 

\subsubsection{Future Work}

In the final subsection of our results, we showed that drivers often utilize coasting and idling during congestion when they are not responding like the parametric models. However, the lack of data forbids us to further investigate and generalize the temporal relations between different clustered behaviors in a continuous congestion event. Future work could be conducted to study the joint temporal distribution or causation between different clusters of behaviors. This may give further insight into how braking and coasting clusters are temporally dependent, providing insight into future modeling of human behavior transitions.

While coasting and idle creep are identified as primary strategies drivers use when models fail to match driver behavior, we are unable to identify specific timing that a driver would initiate or terminate such behavior. We plan, for future research, investigate segments prior and after such lift-and-coast to uncover the stimuli for drivers to initiate such maneuvers. 

\section{Conclusion}

In this study, five kinematic-based parametric Car-Following models are studied under the context of congestion. Their behaviors under different sets of parameters are compared, and the empirical observations for reasons of model-reality difference is discussed. Furthermore, driver behaviors during times when model prediction fails are clustered into five different common maneuvers. The characteristics of the clusters are discussed and their temporal relationship is empirically discussed. 

The performance of all five models on highway-calibrated parameters from various literature consistently fail to capture real driver trajectory, while the calibrated models show noticeable performance in tracking the ground truth trajectory in many scenarios. 
While based on three distinct heuristics, the five models selected for this study produced similar behaviors when calibrated over RMSNE. All five models track trajectory of ego vehicle closely under low ground truth space gap, as theoretically hypothesized. The segments where models all fail to predict human driving behavior is separated and clustered into five categories of maneuvers. The maneuvers are individually discussed, and their temporal relations are empirically observed and discussed. We recommend future modeling of Car-Following behaviors in congested traffic to consider the coasting and idle creep behaviors to better describe real-world driving in congestions. We also recommend future research in distraction recognition and adjacent vehicle intent recognition to better describe human driving patterns in complex and sensitive Car-Following scenarios. 

\section{Acknowledgement}
The authors would like to thank the University of Michigan Transportation Research Institute for supporting the naturalistic driving data. We also thank the support from the Summer Undergraduate Research in Engineering Program of the College of Engineering at the University of Michigan.

\section{Author Contribution Statement}
The authors confirm contribution to the paper as follows: study conception and
design: B. Lin, H. Hou; simulation creation: H. Hou; analysis and
interpretation of results: H. Hou, A. Kusari, B. Lin; draft manuscript
preparation: H. Hou, B. Lin. All authors reviewed the results and approved
the final version of the manuscript.

\newpage

\bibliographystyle{trb}
\bibliography{SURE_citations}

@inproceedings{wu_connections_2018,
	title = {Connections between classical car following models and artificial neural networks},
	url = {https://ieeexplore.ieee.org/document/8569333},
	doi = {10.1109/ITSC.2018.8569333},
	urldate = {2024-06-10},
	booktitle = {2018 21st {International} {Conference} on {Intelligent} {Transportation} {Systems} ({ITSC})},
	author = {Wu, Fangyu and Work, Daniel B.},
	month = nov,
	year = {2018},
	note = {ISSN: 2153-0017},
	pages = {3191--3198},
}

@article{kim_identifying_2023,
	title = {Identifying suitable car-following models to simulate automated vehicles on highways},
	volume = {12},
	issn = {2046-0430},
	url = {https://www.sciencedirect.com/science/article/pii/S2046043023000060},
	doi = {10.1016/j.ijtst.2023.02.003},
	number = {2},
	urldate = {2024-06-11},
	journal = {International Journal of Transportation Science and Technology},
	author = {Kim, Bumsik and Heaslip, Kevin P.},
	month = jun,
	year = {2023},
	pages = {652--664},
}

@article{vasconcelos_calibration_2014,
	series = {17th {Meeting} of the {EURO} {Working} {Group} on {Transportation}, {EWGT2014}, 2-4 {July} 2014, {Sevilla}, {Spain}},
	title = {Calibration of the {Gipps} {Car}-following {Model} {Using} {Trajectory} {Data}},
	volume = {3},
	issn = {2352-1465},
	url = {https://www.sciencedirect.com/science/article/pii/S2352146514002385},
	doi = {10.1016/j.trpro.2014.10.075},
	urldate = {2024-06-11},
	journal = {Transportation Research Procedia},
	author = {Vasconcelos, Luís and Neto, Luís and Santos, Silvia and Silva, Ana Bastos and Seco, Alvaro},
	month = jan,
	year = {2014},
	pages = {952--961},
}

@inproceedings{al-jameel_examining_2009,
	title = {Examining and {Improving} the {Limitations} of {Gazis}-{Herman}-{Rothery} {Car}-following {Model}.},
	author = {Al-jameel, Hamid},
	month = dec,
	year = {2009},
}

@article{zhang_comprehensive_2021,
	title = {A comprehensive comparison study of four classical car-following models based on the large-scale naturalistic driving experiment},
	volume = {113},
	issn = {1569-190X},
	url = {https://www.sciencedirect.com/science/article/pii/S1569190X21000927},
	doi = {10.1016/j.simpat.2021.102383},
	urldate = {2024-06-13},
	journal = {Simulation Modelling Practice and Theory},
	author = {Zhang, Duo and Chen, Xiaoyun and Wang, Junhua and Wang, Yinhai and Sun, Jian},
	month = dec,
	year = {2021},
	pages = {102383},
}

@inproceedings{abbas_analysis_2011,
	title = {Analysis of the {Wiedemann} {Car} {Following} {Model} over {Different} {Speeds} using {Naturalistic} {Data}},
	language = {en},
	author = {Abbas, Montasir M and Medina, Alejandra},
	month = sep,
	year = {2011},
}

@article{treiber_congested_2000,
	title = {Congested traffic states in empirical observations and microscopic simulations},
	volume = {62},
	url = {https://link.aps.org/doi/10.1103/PhysRevE.62.1805},
	doi = {10.1103/PhysRevE.62.1805},
	number = {2},
	urldate = {2024-06-23},
	journal = {Physical Review E},
	author = {Treiber, Martin and Hennecke, Ansgar and Helbing, Dirk},
	month = aug,
	year = {2000},
	note = {Publisher: American Physical Society},
	pages = {1805--1824},
}

@article{zhang_car-following_2024,
	title = {Car-{Following} {Models}: {A} {Multidisciplinary} {Review}},
	issn = {2379-8904},
	shorttitle = {Car-{Following} {Models}},
	url = {https://ieeexplore.ieee.org/abstract/document/10547481?casa_token=KE8JxYPbI7kAAAAA:OsBWY3RTdmeCwVQvh2iVA2mwxQb6lyCinhyzKyCRagECqOTCg_VSRE93PPrVRUMIwCcpJruD6tA},
	doi = {10.1109/TIV.2024.3409468},
	urldate = {2024-06-23},
	journal = {IEEE Transactions on Intelligent Vehicles},
	author = {Zhang, Tianya Terry and Jin, Peter J. and McQuade, Sean T. and Bayen, Alexandre and Piccoli, Benedetto},
	year = {2024},
	note = {Conference Name: IEEE Transactions on Intelligent Vehicles},
	pages = {1--26},
}

@book{treiber_traffic_2013,
	address = {Berlin, Heidelberg},
	title = {Traffic {Flow} {Dynamics}: {Data}, {Models} and {Simulation}},
	copyright = {https://www.springernature.com/gp/researchers/text-and-data-mining},
	isbn = {978-3-642-32459-8 978-3-642-32460-4},
	shorttitle = {Traffic {Flow} {Dynamics}},
	url = {https://link.springer.com/10.1007/978-3-642-32460-4},
	language = {en},
	urldate = {2024-06-24},
	publisher = {Springer Berlin Heidelberg},
	author = {Treiber, Martin and Kesting, Arne},
	year = {2013},
	doi = {10.1007/978-3-642-32460-4},
}

@article{bando_analysis_1998,
	title = {Analysis of optimal velocity model with explicit delay},
	volume = {58},
	copyright = {http://link.aps.org/licenses/aps-default-license},
	issn = {1063-651X, 1095-3787},
	url = {https://link.aps.org/doi/10.1103/PhysRevE.58.5429},
	doi = {10.1103/PhysRevE.58.5429},
	language = {en},
	number = {5},
	urldate = {2024-06-24},
	journal = {Physical Review E},
	author = {Bando, Masako and Hasebe, Katsuya and Nakanishi, Ken and Nakayama, Akihiro},
	month = nov,
	year = {1998},
	pages = {5429--5435},
}

@article{chen_follownet_2023,
	title = {{FollowNet}: {A} {Comprehensive} {Benchmark} for {Car}-{Following} {Behavior} {Modeling}},
	volume = {10},
	copyright = {2023 The Author(s)},
	issn = {2052-4463},
	shorttitle = {{FollowNet}},
	url = {https://www.nature.com/articles/s41597-023-02718-7},
	doi = {10.1038/s41597-023-02718-7},
	language = {en},
	number = {1},
	urldate = {2024-06-25},
	journal = {Scientific Data},
	author = {Chen, Xianda and Zhu, Meixin and Chen, Kehua and Wang, Pengqin and Lu, Hongliang and Zhong, Hui and Han, Xu and Wang, Xuesong and Wang, Yinhai},
	month = nov,
	year = {2023},
	note = {Publisher: Nature Publishing Group},
	pages = {828},
}

@incollection{kerner_traffic_2009,
	address = {New York, NY},
	title = {Traffic {Congestion}, {Modeling} {Approaches} to},
	isbn = {978-0-387-30440-3},
	url = {https://doi.org/10.1007/978-0-387-30440-3_559},
	language = {en},
	urldate = {2024-07-08},
	booktitle = {Encyclopedia of {Complexity} and {Systems} {Science}},
	publisher = {Springer},
	author = {Kerner, Boris S.},
	editor = {Meyers, Robert A.},
	year = {2009},
	doi = {10.1007/978-0-387-30440-3_559},
	pages = {9302--9355},
}

@article{kesting_enhanced_2010,
	title = {Enhanced intelligent driver model to access the impact of driving strategies on traffic capacity},
	volume = {368},
	issn = {1364-503X, 1471-2962},
	url = {https://royalsocietypublishing.org/doi/10.1098/rsta.2010.0084},
	doi = {10.1098/rsta.2010.0084},
	language = {en},
	number = {1928},
	urldate = {2024-07-10},
	journal = {Philosophical Transactions of the Royal Society A: Mathematical, Physical and Engineering Sciences},
	author = {Kesting, Arne and Treiber, Martin and Helbing, Dirk},
	month = oct,
	year = {2010},
	pages = {4585--4605},
}

@article{kesting_calibrating_2008,
	title = {Calibrating {Car}-{Following} {Models} by {Using} {Trajectory} {Data}: {Methodological} {Study}},
	volume = {2088},
	copyright = {http://journals.sagepub.com/page/policies/text-and-data-mining-license},
	issn = {0361-1981, 2169-4052},
	shorttitle = {Calibrating {Car}-{Following} {Models} by {Using} {Trajectory} {Data}},
	url = {http://journals.sagepub.com/doi/10.3141/2088-16},
	doi = {10.3141/2088-16},
	language = {en},
	number = {1},
	urldate = {2024-07-10},
	journal = {Transportation Research Record: Journal of the Transportation Research Board},
	author = {Kesting, Arne and Treiber, Martin},
	month = jan,
	year = {2008},
	pages = {148--156},
}

@article{gad_pygad_2024,
	title = {{PyGAD}: an intuitive genetic algorithm {Python} library},
	volume = {83},
	issn = {1573-7721},
	shorttitle = {{PyGAD}},
	url = {https://doi.org/10.1007/s11042-023-17167-y},
	doi = {10.1007/s11042-023-17167-y},
	language = {en},
	number = {20},
	urldate = {2024-07-13},
	journal = {Multimedia Tools and Applications},
	author = {Gad, Ahmed Fawzy},
	month = jun,
	year = {2024},
	pages = {58029--58042},
}

@article{abdelhalim_real-time_2022,
	title = {A {Real}-{Time} {Safety}-{Based} {Optimal} {Velocity} {Model}},
	volume = {3},
	issn = {2687-7813},
	url = {https://ieeexplore.ieee.org/document/9697070},
	doi = {10.1109/OJITS.2022.3147744},
	urldate = {2024-07-17},
	journal = {IEEE Open Journal of Intelligent Transportation Systems},
	author = {Abdelhalim, Awad and Abbas, Montasir},
	year = {2022},
	note = {Conference Name: IEEE Open Journal of Intelligent Transportation Systems},
	pages = {165--175},
}

@article{bando_dynamical_1995,
	title = {Dynamical model of traffic congestion and numerical simulation},
	volume = {51},
	url = {https://link.aps.org/doi/10.1103/PhysRevE.51.1035},
	doi = {10.1103/PhysRevE.51.1035},
	number = {2},
	urldate = {2024-07-17},
	journal = {Physical Review E},
	author = {Bando, M. and Hasebe, K. and Nakayama, A. and Shibata, A. and Sugiyama, Y.},
	month = feb,
	year = {1995},
	note = {Publisher: American Physical Society},
	pages = {1035--1042},
}

@techreport{sayer_integrated_2011,
	title = {Integrated {Vehicle}-{Based} {Safety} {Systems} {Field} {Operational} {Test} : {Final} {Program} {Report}},
	shorttitle = {Integrated {Vehicle}-{Based} {Safety} {Systems} {Field} {Operational} {Test}},
	url = {https://rosap.ntl.bts.gov/view/dot/3358},
	language = {English},
	number = {FHWA-JPO-11-150;UMTRI-2010-36},
	urldate = {2024-07-23},
	institution = {University of Michigan},
	author = {Sayer, J. and LeBlanc, D. and Bogard, S. and Funkhouser, D. and Bao, S. and Buonarosa, M. L. and Blankespoor, A. and {University of Michigan. Transportation Research Institute}},
	month = jun,
	year = {2011},
}

@article{gipps_queueing_1977,
	title = {A {Queueing} {Model} for {Traffic} {Flow}},
	volume = {39},
	issn = {0035-9246},
	url = {https://doi.org/10.1111/j.2517-6161.1977.tb01626.x},
	doi = {10.1111/j.2517-6161.1977.tb01626.x},
	number = {2},
	urldate = {2024-07-23},
	journal = {Journal of the Royal Statistical Society: Series B (Methodological)},
	author = {Gipps, P. G.},
	month = jan,
	year = {1977},
	pages = {276--282},
}

@article{jiang_full_2001,
	title = {Full velocity difference model for a car-following theory},
	volume = {64},
	url = {https://link.aps.org/doi/10.1103/PhysRevE.64.017101},
	doi = {10.1103/PhysRevE.64.017101},
	number = {1},
	urldate = {2024-07-23},
	journal = {Physical Review E},
	author = {Jiang, Rui and Wu, Qingsong and Zhu, Zuojin},
	month = jun,
	year = {2001},
	note = {Publisher: American Physical Society},
	pages = {017101},
}

@inproceedings{krajewski_highd_2018,
	address = {Maui, HI, USA},
	title = {The {highD} {Dataset}: {A} {Drone} {Dataset} of {Naturalistic} {Vehicle} {Trajectories} on {German} {Highways} for {Validation} of {Highly} {Automated} {Driving} {Systems}},
	isbn = {978-1-72810-321-1},
	shorttitle = {The {highD} {Dataset}},
	url = {https://doi.org/10.1109/ITSC.2018.8569552},
	doi = {10.1109/ITSC.2018.8569552},
	urldate = {2024-07-26},
	booktitle = {2018 21st {International} {Conference} on {Intelligent} {Transportation} {Systems} ({ITSC})},
	publisher = {IEEE Press},
	author = {Krajewski, Robert and Bock, Julian and Kloeker, Laurent and Eckstein, Lutz},
	month = nov,
	year = {2018},
	pages = {2118--2125},
}

@misc{paszke_pytorch_2019,
	title = {{PyTorch}: {An} {Imperative} {Style}, {High}-{Performance} {Deep} {Learning} {Library}},
	shorttitle = {{PyTorch}},
	url = {http://arxiv.org/abs/1912.01703},
	doi = {10.48550/arXiv.1912.01703},
	urldate = {2024-07-27},
	publisher = {arXiv},
	author = {Paszke, Adam and Gross, Sam and Massa, Francisco and Lerer, Adam and Bradbury, James and Chanan, Gregory and Killeen, Trevor and Lin, Zeming and Gimelshein, Natalia and Antiga, Luca and Desmaison, Alban and Köpf, Andreas and Yang, Edward and DeVito, Zach and Raison, Martin and Tejani, Alykhan and Chilamkurthy, Sasank and Steiner, Benoit and Fang, Lu and Bai, Junjie and Chintala, Soumith},
	month = dec,
	year = {2019},
	note = {arXiv:1912.01703 [cs, stat]},
}

@article{chandler_traffic_1958,
	title = {Traffic {Dynamics}: {Studies} in {Car} {Following}},
	volume = {6},
	issn = {0030-364X},
	shorttitle = {Traffic {Dynamics}},
	url = {https://pubsonline.informs.org/doi/abs/10.1287/opre.6.2.165},
	doi = {10.1287/opre.6.2.165},
	number = {2},
	urldate = {2024-07-29},
	journal = {Operations Research},
	author = {Chandler, Robert E. and Herman, Robert and Montroll, Elliott W.},
	month = apr,
	year = {1958},
	note = {Publisher: INFORMS},
	pages = {165--184},
}

@article{taylor_method_2015,
	title = {Method for investigating intradriver heterogeneity using vehicle trajectory data: {A} {Dynamic} {Time} {Warping} approach},
	volume = {73},
	issn = {0191-2615},
	shorttitle = {Method for investigating intradriver heterogeneity using vehicle trajectory data},
	url = {https://www.sciencedirect.com/science/article/pii/S0191261514002264},
	doi = {10.1016/j.trb.2014.12.009},
	urldate = {2024-07-30},
	journal = {Transportation Research Part B: Methodological},
	author = {Taylor, Jeffrey and Zhou, Xuesong and Rouphail, Nagui M. and Porter, Richard J.},
	month = mar,
	year = {2015},
	pages = {59--80},
}

@article{tavenard_tslearn_2020,
	title = {Tslearn, {A} {Machine} {Learning} {Toolkit} for {Time} {Series} {Data}},
	author = {Tavenard, Romain and Faouzi, Johann and Vandewiele, Gilles and Divo, Felix and Androz, Guillaume and Holtz, Chester and Payne, Marie and Yurchak, Roman and Rußwurm, Marc and Kolar, Kushal and Woods, Eli},
	month = jan,
	year = {2020},
}

@article{wiedemann_simulation_1974,
	title = {{SIMULATION} {DES} {STRASSENVERKEHRSFLUSSES}.},
	url = {https://trid.trb.org/View/596235},
	urldate = {2024-07-31},
	author = {Wiedemann, Rainer},
	year = {1974},
}

@inproceedings{timbario_testing_2022,
	title = {Testing and {Modeling} of {Engine} {Idle} {Creep} in {Conventional} {Automatic}-{Equipped} {Vehicles}},
	language = {en},
	publisher = {EDC Corporation},
	author = {Timbario, Thomas A and Sheldon, Stuart and Stoner, Jacob and Nelson, Jonathan D},
	month = mar,
	year = {2022},
}

@techreport{timbario_engine_2023,
	address = {Warrendale, PA},
	type = {{SAE} {Technical} {Paper}},
	title = {Engine {Idle} {Creep} {Testing} and {Modeling} of {Vehicles} {Equipped} with {CVT}, {DCT}, and {Conventional} {Automatic} {Transmissions}},
	url = {https://www.sae.org/publications/technical-papers/content/2023-01-0620/},
	language = {English},
	number = {2023-01-0620},
	urldate = {2024-08-01},
	institution = {SAE Technical Paper},
	author = {Timbario, Thomas A. and Stoner, Jacob and Ii, Stuart Sheldon},
	month = apr,
	year = {2023},
	doi = {10.4271/2023-01-0620},
	note = {ISSN: 0148-7191, 2688-3627},
}

@article{kim_asymmetric_2024,
	title = {Asymmetric repulsive force model: {A} new car-following model with psycho-physical characteristics},
	volume = {161},
	issn = {0968-090X},
	shorttitle = {Asymmetric repulsive force model},
	url = {https://www.sciencedirect.com/science/article/pii/S0968090X24000925},
	doi = {10.1016/j.trc.2024.104571},
	urldate = {2024-08-01},
	journal = {Transportation Research Part C: Emerging Technologies},
	author = {Kim, Yeeun and Yeo, Hwasoo},
	month = apr,
	year = {2024},
	pages = {104571},
}

@article{campbell_safety_2012,
	title = {{SAFETY} {The} {SHRP} 2 {Naturalistic} {Driving} {Study}},
	language = {en},
	author = {Campbell, Kenneth L},
	year = {2012},
}

\end{document}